\documentclass[%
 reprint,
 amsmath,amssymb,
 aps,
prd,
floatfix,
]{revtex4-2}

\usepackage{placeins}
\usepackage{xcolor}
\usepackage{graphicx}% Include figure files
\usepackage{dcolumn}% Align table columns on decimal point
\usepackage{bm}% bold math
\usepackage{hyperref}% add hypertext capabilities

\usepackage{amsmath}
\usepackage{amssymb}
\usepackage{amsfonts}
\usepackage{accents}
\usepackage{algorithm}
\usepackage{algpseudocode}
\usepackage{float}

\newcommand*{\Scale}[2][4]{\scalebox{#1}{$#2$}}



\newboolean{usePNG}
\setboolean{usePNG}{true}

\def\g{\gamma}

\def\D{\Delta}
\def\p{\partial}

\def\lm{\mathcal{L}_m}

\def\lb{\mathcal{L}_\beta}

\def\n{\nabla}
\def\e4c{e^{-4\chi}}
\def\ep4c{e^{4\chi}}

\def\gh{\hat{\gamma}}

\def\Rh{\hat{R}}
\def\Ah{\hat{A}}
\def\Dh{\hat{\Delta}}

\def\zh{\bar{Z}}
\def\Zh{\hat{Z}}
\def\Kh{\hat{K}}
\def\alphah{\hat{\alpha}}
\def\betah{\hat{\beta}}
\def\Bh{\hat{B}}
\def\chih{\hat{\chi}}
\def\ah{\hat{a}}
\def\Xh{\hat{X}}
\def\bh{\hat{b}}
\def\lh{\hat{l}}
\def\LLh{\hat{L}}
\def\uh{\hat{u}}

\def\Yh{\hat{Y}}

\def\Omegah{\hat{\Omega}}

\def\Xhb{\hat{X}^{'}}
\def\Yhb{\hat{Y}^{'}}
\def\lhb{\hat{l}^{'}}
\def\bhb{\hat{b}^{'}}
\def\Bhb{\hat{B}^{'}}
\def\LLhb{\hat{L}^{'}}

\def\Zhb{\hat{Z}^{'}}

\def\gth{\hat{\accentset{\sim}{\gamma}}}
\def\Lth{\hat{\accentset{\sim}{\Lambda}}}

\def\Ath{\hat{\accentset{\sim}{A}}}
\def\Dth{\hat{\accentset{\sim}{\Delta}}}
\def\Lamt{\accentset{\sim}{\Lambda}}

\def\wabs{\left|\omega\right|}
\def\omegat{\accentset{\sim}{\omega}}

\def\Thetat{\accentset{\sim}{\Theta}}

\def\no{\accentset{\circ}{\nabla}}
\def\Do{\accentset{\circ}{D}}
\def\go{\accentset{\hphantom{i}\circ}{\g}}

\def\Go{\accentset{\circ}{\Gamma}}
\def\Ro{\accentset{\circ}{R}}

\def\zo{\accentset{\circ}Z}

\def\gt{\accentset{\hphantom{i}\sim}{\g}}
\def\Gt{\accentset{\sim}{\Gamma}}

\def\dt{\accentset{\sim}{D}}
\def\Rt{\accentset{\sim}{R}}

\def\At{\accentset{\sim}{A}}
\def\Dt{\accentset{\sim}{\D}}
\def\jt{\accentset{\sim}{j}}

\def\lp{\left}
\def\rp{\right}

\def\tf{\mathrm{TF}}
\def\vp{\vphantom{}}

\def\lnalpha{\ln{\lp(\alpha\rp)}}

\begin{document}

\preprint{APS/123-QED}

\title{RCCZ4: A Reference Metric Approach to Z4}

\author{Gray D. Reid}
\affiliation{Department of Physics and Astronomy,
     University of British Columbia,
     Vancouver BC, V6T 1Z1 Canada}

\author{Matthew W. Choptuik}
\affiliation{
     Department of Physics and Astronomy,
     University of British Columbia,
     Vancouver BC, V6T 1Z1 Canada}

\date{\today}% It is always \today, today,

\begin{abstract}

The hyperbolic formulations of numerical relativity due to Baumgarte, 
Shapiro, Shibata \& Nakamura (BSSN) and Nagy Ortiz \& Reula (NOR), among 
others, achieve stability through the effective embedding of general 
relativity within the larger Z4 system. In doing so, various elliptic 
constraints  are promoted to dynamical degrees of freedom, permitting the 
advection of constraint violating modes. Here we demonstrate that it is 
possible to achieve equivalent performance through a modification of 
fully covariant and conformal Z4 (FCCZ4) wherein constraint violations 
are coupled to a reference metric completely independently of the physical 
metric. We show that this approach works in the presence of black holes and 
holds up robustly in a variety of spherically symmetric simulations 
including the critical collapse of a scalar field. We then demonstrate that 
our formulation is strongly hyperbolic through the use of a 
pseudodifferential first order reduction and compare its 
hyperbolicity properties to those of FCCZ4 and 
generalized BSSN (GBSSN).

Our present approach makes use of a static Lorentzian reference metric and 
does not appear to provide significant advantages over FCCZ4. However, we 
speculate that dynamical specification of the reference metric may provide a 
means of exerting greater control over constraint violations than what is 
provided by current BSSN-type formulations.

\end{abstract}

\maketitle

\section{\label{rccz4_sec_introduction}Introduction}
The formulations of numerical relativity based on the Baumgarte, Shapiro, 
Shibata \& Nakamura (BSSN) decomposition effectively achieve strong 
hyperbolicity and stability by performing a partial embedding of general 
relativity (GR) within 
the larger Z4 system~\cite{alcubierre2011formulations, sanchis2014fully, 
hilditch2013compact}. In this paper 
we demonstrate that the Z4 system is not uniquely suitable for this purpose 
and present an alternative formulation of GR that is also well suited for 
numerical relativity. This formulation is based on an alternative embedding 
of GR and holds up well in a variety of simulations in spherical symmetry 
including those of black holes with puncture initial data as well as in the 
critical collapse of the massless scalar field. Additionally, we show that 
our new formulation is strongly hyperbolic and, in fact, has the same 
principal symbol as fully covariant and conformal Z4 (FCCZ4).

The Z4 formulation takes its name from the introduction of a four 
vector, $Z_\mu$, to the Einstein equations, 
\begin{align}
   \label{rccz4_Z4_v1}
   R_{\mu\nu} + 2\n_{(\mu}Z_{\nu)}
   - 8\pi \lp( T_{\mu\nu} -\frac{1}{2}g_{\mu\nu}T \rp) &= 0.
\end{align}
In the context of general relativity, the evolution of this system acts to 
advect and/or damp violations of the Hamiltonian and momentum constraints. 
In the limit $Z_{\mu} \rightarrow 0$ we recover GR~\cite{bona2003general}.

If we examine formulations such as NOR~\cite{nagy2004strongly} and generalized 
BSSN (GBSSN)~\cite{brown2009covariant} in detail, we find that they are 
essentially minor variations of Z4-derivable formulations in which the 
temporal component of $Z_{\mu}$ is not evolved and substitutions or additions 
of the Hamiltonian and momentum constraints have been 
made~\cite{alcubierre2011formulations, sanchis2014fully, daverio2018apples, 
hilditch2013compact, alic2012conformal, alic2013constraint}. 
The case could also be made that the equations of motion of Z4 formalisms 
arise naturally while those of NOR and GBSSN come from experimentation 
to achieve stability and strong hyperbolicity.

In that same spirit of experimentation, we note that if we assume the Einstein 
equations are very nearly satisfied, such that their violation is contained 
in a tensor, $E_{\mu\nu}$:
\begin{align}
   \epsilon E_{\mu\nu} & =8\pi\lp(T_{\mu\nu} -\frac{1}{2}g_{\mu\nu}T\rp) 
   - R_{\mu\nu},
\end{align}
where $\epsilon \ll 1$, then
the Z4 equations (\ref{rccz4_Z4_v1}) may be written as
\begin{align}
\label{rccz4_n_u_Z_v_p_n_v_Z_u}
   \n_\mu Z_\nu + \n_\nu Z_\mu &= \epsilon E_{\mu \nu},
\end{align}
with trace given by
\begin{align}
   \label{rccz4_n_u_Z_u}
   \n_\mu Z^\mu &= \frac{1}{2}\epsilon E_\mu\vp^{\mu}.
\end{align}
Taking the divergence of (\ref{rccz4_n_u_Z_v_p_n_v_Z_u}) and using the 
commutator of covariant derivatives, we find
\begin{align}
   \Box Z^\nu  &= - \n_\mu \n^\nu Z^\mu + \n_\mu \epsilon E^{\mu \nu},
\\ \nonumber
   &= - \n^\nu \n_\mu Z^\mu 
   + R^\mu\vp_\alpha\vp^\nu\vp_\mu Z^\alpha + \epsilon \n_\mu E^{\mu \nu},
\\ \nonumber
   &= \epsilon \lp( - \frac{1}{2} \n^\nu {E_\mu}^\mu 
   + \n_\mu E^{\mu \nu} \rp) - R_\mu\vp^\nu Z^\mu. 
\end{align}

Heuristically, $Z_\mu$ evolves according to some complicated wave equation on
$g_{\mu\nu}$, which is sourced by the deviation from the Einstein equations. 
This is desirable since it means that $Z_\mu$ has characteristics with 
magnitude $\sim 1$ on $g_{\mu\nu}$ when $\epsilon$ is small and $g_{\mu\nu}$ 
is not too curved. 
In the presence of significant curvature, however, the picture is less clear 
and we note that we have completely ignored the backreaction of $Z_{\mu}$ 
on $E_{\mu\nu}$.

If we modify the Z4 formulation such that $Z_{\mu}$ is no longer directly
coupled to the physical metric, and is instead coupled to some other metric 
$\accentset{\circ}{g}_{\mu\nu}$ with associated connection $\no_{\mu}$:
\begin{align}
\label{rccz4_RZ4_equation}
   \no_\mu Z_\nu + \no_\nu Z_\mu &= 8\pi\lp(T_{\mu\nu} 
   -\frac{1}{2}g_{\mu\nu}T\rp) - R_{\mu\nu},
\end{align}
we find,
\begin{align}
\label{rccz4_no_u_Zo_v_p_no_v_Z_u}
   \no_\mu Z_\nu + \no_\nu Z_\mu &= \epsilon {E}_{\mu\nu},
\end{align}
with trace:
\begin{align}
\label{rccz4_no_u_Zo_u}
   \no_\mu \zo^\mu &= \frac{1}{2}\epsilon\accentset{\circ}{E}_\mu\vp^\mu.
\end{align}
Here, variables accented with ``$\circ$" have had a 
covariant tensorial index raised with 
$\accentset{\circ}{g}^{\mu\nu}$. 
Taking the divergence of (\ref{rccz4_no_u_Zo_v_p_no_v_Z_u}),  we find:
\begin{align}
   \accentset{\circ}{\Box} \zo^\nu  &= 
   - \no_\mu \no^\nu \zo^\mu + \no_\mu \epsilon 
   \accentset{\circ}{E}^{\mu \nu},
\\ \nonumber
   &= 
   - \no^\nu \no_\mu \zo^\mu 
   + \Ro^\mu\vp_\alpha\vp^\nu\vp_\mu Z^\alpha + \epsilon \no_\mu 
   \accentset{\circ}{E}^{\mu \nu},
\\ \nonumber
   &= 
   \epsilon \lp( - \frac{1}{2} \no^\nu {\accentset{\circ}{E}_\mu}^\mu 
   + \no_\mu \accentset{\circ}{E}^{\mu \nu} \rp) 
   - \Ro_\mu\vp^\nu \zo^\mu. 
\end{align}
As such, if we choose $\accentset{\circ}{g}_{\mu\nu}$ so that 
$\Ro_{\mu\nu}$ vanishes, we might expect $\zo^{\mu}$ to propagate with speed 
$\sim 1$ on $\accentset{\circ}{g}_{\mu\nu}$ when $\epsilon$ is small,
regardless of the curvature of $g_{\mu\nu}$. Although 
Sec.~\ref{rccz4_sec_hyperbolicity} 
demonstrates that this intuition does not hold in practice, it served to 
motivate the original investigation and
\textcolor{black}{
the core concept bears some resemblance to the modified harmonic 
gauges of Kovacs and Reall in which an auxiliary metric is used to control 
the speed of propagation of constraint violating 
modes~\cite{kovacs2020well1, kovacs2020well2}.}
In what follows, we expand upon this idea and
present a formulation of the Einstein equations based on a flat, 
time-invariant reference metric which yields a system which performs very 
similarly to the standard GBSSN~\cite{brown2009covariant, 
alcubierre2011formulations} and FCCZ4~\cite{sanchis2014fully} formulations. 
Further work with dynamical specification of the reference metric may allow 
for more fine-grained control over constraint damping and stability 
properties. 

In Sec.~\ref{rccz4_sec_derivation} we give a brief derivation of our 
formulation; a more detailed derivation may be found in  
Appendices~\ref{app_rccz4_rz4_derivation} and \ref{app_rccz4_rccz4_derivation}. 
Section~\ref{rccz4_sec_GBSSN_FCCZ4_EoM} introduces the equations of motion 
for the GBSSN and FCCZ4 formulations of numerical relativity which we make 
use of in our various comparative analyses. In 
Sec.~\ref{rccz4_sec_comparison} we compare the performance of our 
formulation with FCCZ4 and GBSSN in a variety of numerical tests including 
strong field convergence testing, simulation of black holes and the critical 
collapse of the scalar field in spherical symmetry. After demonstrating that 
the method works in spherical symmetry, we shift gears and analyse the 
hyperbolicity of our approach:~Sec.~\ref{rccz4_sec_hyperbolicity} sees us 
derive the conditions under which our method is strongly hyperbolic and 
examine how it compares to both GBSSN and FCCZ4. Finally, in 
Sec.~\ref{rccz4_sec_conclusions} we present our conclusions and suggestions 
for future research into related formulations of numerical relativity.

\section{\label{rccz4_sec_derivation}Derivation of RCCZ4}

We begin with the Z4 equations coupled to a reference metric as 
in~(\ref{rccz4_RZ4_equation}), which we refer to as \em reference metric 
Z4 \em (RZ4), with the aim of developing an ADM decomposition equivalent of 
the system. Once we have this initial value formulation, we perform a 
decomposition similar to GBSSN or FCCZ4 in terms of a conformal metric and 
conformal trace-free extrinsic curvature, arriving at \em reference metric 
covariant and conformal Z4 \em (RCCZ4). Again, more details are provided in 
Appendices~\ref{app_rccz4_rz4_derivation} and \ref{app_rccz4_rccz4_derivation}.

Using standard notation in which $n^{\mu}$ is the unit normal to the 
foliation in a {3+1} decomposition, $\alpha$ is the lapse, $\beta^{i}$ is 
the shift and $\g_{ij}$ is the induced 3-metric on the foliation,  the RZ4 
equations (with damping parameters $\kappa_1$ and $\kappa_2$) may be written 
in canonical form as:
\begin{align}
   \label{rccz4_RZ4_v2}
   &R_{\mu\nu} -\frac{1}{2}g_{\mu\nu} R + 2\no_{(\mu}Z_{\nu)} 
   -g_{\mu\nu}\no_{(\alpha} Z_{\beta)} g^{\alpha\beta} 
   \\ \nonumber
   & \hphantom{=}
   -\kappa_1 \lp[ \frac{}{} 2 n_{(\mu} Z_{\nu)} + \kappa_2 g_{\mu\nu} 
   n_{\sigma} Z^{\sigma} \rp] - 8\pi T_{\mu\nu} =0 .
\end{align}
Equivalently, the trace reversed form is:
\begin{align}
   \label{rccz4_RZ4_v1}
   &R_{\mu\nu} + 2\no_{(\mu}Z_{\nu)}
   - 8\pi \lp( T_{\mu\nu}  -\frac{1}{2}g_{\mu\nu}T \rp) 
   \\ \nonumber
   & \hphantom{=}
   -\kappa_1 \lp[ \frac{}{} 2 
   n_{(\mu} Z_{\nu)} - \lp(1 + \kappa_2\rp) g_{\mu\nu} n_{\sigma} 
   Z^{\sigma} \rp]  = 0.
\end{align}
Taking the trace (with respect to $g^{\mu\nu}$) of 
(\ref{rccz4_RZ4_v1}) yields:
\begin{align}
   \label{rccz4_RZ4_trace}
   &R + 2\no_{(\mu} Z_{\nu)}g^{\mu\nu} +\kappa_1 \lp(2+4\kappa_2\rp) 
   n_{\mu} Z^{\mu} 
   \\ \nonumber 
   & \hphantom{=}
   +8\pi T =0.
\end{align}

From here we roughly follow the ADM derivations 
of~\cite{alcubierre2008introduction, gourgoulhon2012} and take projections of 
(\ref{rccz4_RZ4_v2})--(\ref{rccz4_RZ4_trace}) onto and orthogonal to the 
spatial hypersurfaces which foliate four dimensional spacetime in a standard 
{3+1} decomposition 
(see Appendix~\ref{app_rccz4_rz4_derivation}).
\textcolor{black}{
As the 
focus of this  paper is the exploration of the feasibility of  
alternative embeddings of general relativity, we have made the 
choice to simplify our investigation and forgo all forms of scale dependent 
damping. In what follows, we set $\kappa_1 = \kappa_2 = 0$ 
in~(\ref{rccz4_RZ4_v2})--(\ref{rccz4_RZ4_trace}) yielding the simpler set 
of equations:}
\begin{align}
   \label{rccz4_RZ4_v2_simp}
   &R_{\mu\nu} -\frac{1}{2}g_{\mu\nu} R + 2\no_{(\mu}Z_{\nu)} 
   -g_{\mu\nu}\no_{(\alpha} Z_{\beta)} g^{\alpha\beta} 
   \\ \nonumber
   &
   -8\pi T_{\mu\nu}= 0,
\\
   \label{rccz4_RZ4_v1_simp}
   &R_{\mu\nu} + 2\no_{(\mu}Z_{\nu)}
   - 8\pi \lp( T_{\mu\nu}  -\frac{1}{2}g_{\mu\nu}T \rp) =0,
\\
   \label{rccz4_RZ4_trace_simp}
   &R + 2\no_{(\mu} Z_{\nu)}g^{\mu\nu} +8\pi T =0.
\end{align}

We have considered only the simplest case where $\accentset{\circ}{g}_{\mu\nu}$ 
is a time-invariant, curvature-free Lorentzian metric with 
$\accentset{\circ}{g}_{tt} = -1, \accentset{\circ}{g}_{tj} = 0$. With these 
restrictions,
projection of (\ref{rccz4_RZ4_v2_simp})--(\ref{rccz4_RZ4_trace_simp}) yields 
the ADM equivalent of the RZ4 equations:
\begin{align}
\label{rccz4_gamma_ij}
        \lm\g_{ij} &= -2\alpha K_{ij},    
\\
\label{rccz4_k_ij}
        \lm K_{ij} &= -D_i D_j \alpha +\alpha \lp( R_{ij} + K K_{ij} 
        -2 K_{ik} K^{k}\vp_{j}\rp)
        \\ \nonumber
        & \hphantom{=}
        +4\pi\alpha \lp( \lp[ S-\rho\rp]\g_{ij} -2S_{ij} \rp) + 2 \alpha
        \Do_{(i} \zh_{j)},
\\
\label{rccz4_theta}
        \lm \Theta &= \frac{\alpha}{2}\lp(R + K^2 - K_{ij}K^{ij}
        - 16\pi\rho \rp) 
        \\ \nonumber
        & \hphantom{=}
        + \alpha\g^{ij}\Do_i \zh_{j} -\frac{\Theta}{\alpha}\lm \alpha 
        \\ \nonumber
        & \hphantom{=}
        +\frac{\zh_i}{\alpha}\lp(\lm \beta^i - \beta^j\Do_j\beta^i\rp),
\\
\label{rccz4_z_i}        
        \lm \zh_i &= \alpha \lp( D_j K^{j}\vp_i -D_i K -8\pi j_i \rp) 
        -2 \zh_j\Do_i\beta^j 
        \\ \nonumber
        & \hphantom{=}
        +\Theta\Do_i\alpha +\alpha\Do_i\Theta,
\end{align} 
where $\lm = \p_t - \lb$ and the quantities $\Theta$ and $\zh_i$ are defined as,
\begin{align}
        \Theta &= -n_\mu Z^\mu,
\\
        \zh_i &= \g^{\mu}\vp_i Z_{\mu},
\\
        \zh^i &= \g^{ij} \zh_j.
\end{align}
Once again, we direct readers to Appendix~\ref{app_rccz4_rz4_derivation}
for a more detailed derivation.

In order to cast (\ref{rccz4_gamma_ij})--(\ref{rccz4_z_i}) in a form better 
suited to evolving generic spacetimes, we perform the same covariant and 
conformal decomposition that we would for GBSSN and FCCZ4. We rewrite the 
3-metric, $\g_{ij}$, and extrinsic curvature, $K_{ij}$, in terms of the 
conformal factor, $\chi$, the conformal metric, $\gt_{ij}$, the trace of the 
extrinsic curvature, $K$, and the trace-free extrinsic curvature $\At_{ij}$:
\begin{align}
\label{rccz4_g_def}
   \g_{ij} &= e^{4\chi}\gt_{ij},
\\
\label{rccz4_K_def}
   K_{ij} &= \ep4c\lp( \At_{ij} - \frac{1}{3} \gt_{ij} K \rp).
\end{align}
We also define the quantities ${\Dt^{i}}_{jk}$ and $\Dt^{i}$ in terms of the 
difference between the Christoffel symbols of $\gt_{ij}$ and those of a flat 
background 3-metric $\go_{ij}$: the latter is chosen to coincide with 
the spatial portion of $\go_{\mu\nu}$:
\begin{align}
   \Gt^i &= \Gt^{i}\vp_{j k} \gt^{jk},
\\
   {\Dt^{i}}_{jk} &= {\Gt^{i}}_{jk} - {\Go^{i}_{jk}},
\\
   \Dt^i &= \Gt^i - \Go^{i}\vp_{jk} \gt^{jk}.
\end{align}
Additionally, we define
the quantity $\Lamt^{i}$ which plays the same role as the 
conformal connection functions in BSSN~\cite{alcubierre2011formulations} 
and FCCZ4~\cite{sanchis2014fully}:
\begin{align}
\label{rccz4_Lamt_def}
   \Lamt^i &= \Dt^i + 2\gt^{ij} \zh_j,
\\
\label{rccz4_Z_def}
   \zh^{i} &= \frac{1}{2} \lp( \Lamt^i - \Dt^i\rp)\e4c.
\end{align}
Finally, adopting the Lagrangian choice for the evolution  of the 
determinant of the conformal metric: 
\begin{align}
   \p_t \gt &= 0,
\end{align}
and defining the quantity $\Thetat$ in terms of $\Theta$, $\alpha$, $\zh_i$ 
and $\beta^i$:
\begin{align}
\label{rccz4_Theta2_rccz4}
   \Thetat &= \alpha\Theta - \beta^i \zh_i,
\end{align}
we find the RCCZ4 equations of motion:
\begin{align}
\label{rccz4_chi_rccz4}
        \lm \chi &= -\frac{1}{6} \alpha K + \frac{1}{6}  \dt_m\beta^m,
\\
\label{rccz4_K_rccz4}
        \lm K &= -D^2\alpha +\alpha \lp( R +K^2
        +2\g^{ij} \Do_{(i} \zh_{j)} 
        \rp.
        \\ \nonumber
        &\mathopen{}\lp.\vphantom{\frac{}{}} \hphantom{=}
        +4 \pi \lp(S-3\rho\rp) \rp),
\\
\label{rccz4_Theta_rccz4}
        \lm \Thetat &= \frac{\alpha^2}{2}\lp(R -\At_{ij} \At^{ij}
        +\frac{2}{3} K^2 -16\pi\rho 
        \rp.
        \\ \nonumber
        &\mathopen{}\lp.\vphantom{\frac{}{}} \hphantom{=}
        + 2\g^{ij}\Do_i \zh_j \rp) -\beta^{j}\lp( \beta^l\Do_j \zh_l + 
        \Do_j\Thetat \rp)
        \\ \nonumber
        & \hphantom{=}
        -\alpha\beta^{j} \lp( D_l {\At^{l}}_{j} -\frac{2}{3} \dt_j K 
        -8\pi j_j \rp),
\\
\label{rccz4_gh_rccz4}
        \lm \gt_{ij} &= -2 \alpha \At_{ij} - \frac{2}{3} \gt_{ij} \dt_{m}
        \beta^{m},
\\
\label{rccz4_ah_rccz4}
        \lm \At_{ij} &= \e4c\lp[ -D_i D_j \alpha +\alpha R_{ij} 
        - 8\pi\alpha S_{ij} 
        \rp.
        \\ \nonumber
        &\mathopen{}\lp. \hphantom{=}
        +2\alpha\Do_{(i} \zh_{j)} \rp]^\tf +\alpha\lp(K \At_{ij} 
        -2 \At_{ik} {\At^{k}}_j\rp) 
        \\ \nonumber
        & \hphantom{=}
        -\frac{2}{3}
        \At_{ij}\dt_l \beta^{l},
\\
\label{rccz4_Lh_rccz4}
        \lm \Lamt^i &= \gt^{mn} \Do_m \Do_n \beta^i -2\At^{im}\dt_m\alpha
        \\ \nonumber
        & \hphantom{=}
        +2\alpha\At^{mn}\Dt^{i}\vp_{mn} +\frac{1}{3}\dt^{i}\dt_n\beta^n
        +\frac{2}{3} \Lamt^{i}\dt_n \beta^n
        \\ \nonumber
        & \hphantom{=}
        + 4\alpha \lp( \zh_j\At^{ij} 
        + 3 \At^{li}\dt_l\chi -\frac{1}{3}\dt^i K -4\pi\jt^i \rp)
        \\ \nonumber
        & \hphantom{=}
        +2\dt^i \Thetat + 2\gt^{ij}\lp( \beta^l\ \Do_{j}\zh_l 
        - \zh_l \Do_j \beta^l \rp),
\\
\label{rccz4_Zh_rccz4}
       \lm \zh_i &= \alpha\lp[ D_l{\At^{l}}_i -\frac{2}{3}\dt_i K 
       -8\pi j_i \rp] - \zh_l\Do_i\beta^l 
       \\ \nonumber
       & \hphantom{=}
       + \beta^{l}\Do_i \zh_{l} + \Do_i\Thetat \, .
\end{align}
Here, either $\zh_{i}$ or $\Lamt^i$ may be viewed as the dynamical quantity
associated with the momentum constraint violations and all quantities 
denoted by a tilde are raised and lowered with the conformal metric. 
``$\mathrm{TF}$" denotes trace free with respect to the 3-metric $\g_{ij}$ 
and the Ricci tensor may be split into scale-factor and conformal parts as
\begin{align}
\label{rccz4_R_def}
   R_{ij} &= \Rt_{ij} + R^{\chi}_{ij},
\end{align}
with
\begin{align}
\label{rccz4_R_tilde_def}
   \Rt_{ij} &= -\frac{1}{2}\gt^{mn}\Do_m\Do_n\gt_{ij} + \gt_{m(i}\Do_{j)}
   \Dt^m 
   \\ \nonumber
   & \hphantom{=}
   + \Dt^m\Dt_{(ij)m} +2\Dt^{mn}\vp_{(i}\Dt_{j)mn} + \Dt^{mn}\vp_{i}
   \Dt_{mnj},
\\
\label{rccz4_R_chi_def}
   R^{\chi}_{ij} &= -2 \dt_i\dt_j \chi - 2\gt_{ij}\dt^{k}\dt_k \chi 
   + 4\dt_i\chi\dt_j \chi 
   \\ \nonumber
   & \hphantom{=}
   - 4 \gt_{ij} \dt^k \chi\dt_k \chi.
\end{align}

Note that the equations of motion for $\Thetat$, (\ref{rccz4_Theta_rccz4}), 
and $\zh_i$, (\ref{rccz4_Zh_rccz4}), are essentially sourced by violations of 
the Hamiltonian and momentum constraints respectively. In terms of the 
conformal decomposition these constraints then take the form
\begin{align}
   H &= \frac{1}{2} \lp( R + \frac{2}{3} K^2 - \At_{ij}\At^{ij} \rp)
   - 8 \pi \rho,
\\
   M^i &= \e4c\lp( \dt_j \At^{ij} - \frac{2}{3} \gt^{ij}\dt_j K 
   + 6 \At^{ij}\dt_{j} \chi 
   \rp.
   \\ \nonumber
   &\mathopen{}\lp. \hphantom{=}
   - 8\pi \jt^i \rp).
\end{align}

\section{FCCZ4 and GBSSN Equations of Motion
\label{rccz4_sec_GBSSN_FCCZ4_EoM}}

In testing the viability of RCCZ4 as a formulation for numerical relativity, 
we make use of the formulation of FCCZ4 due to Sanchis-Gual et 
al.~\cite{sanchis2014fully} along with the formulation of GBSSN by 
Brown~\cite{brown2008bssn} as presented by Alcubierre and 
Mendaz~\cite{alcubierre2011formulations}. In our notation, the equations of 
motion for FCCZ4 are:
\begin{align}
\label{rccz4_chi_fccz4}
        \lm \chi &= -\frac{1}{6} \alpha K + \frac{1}{6}  \dt_m\beta^m,
\\
\label{rccz4_K_fccz4}
        \lm K &= -D^2\alpha +\alpha R +\alpha \lp(K^2-2\Theta K\rp)
        \\ \nonumber
        & \hphantom{=}
        +2\alpha D_i \zh^i +4 \pi \alpha\lp(S-3\rho\rp),
\\
\label{rccz4_Theta}
        \lm \Theta &= \frac{\alpha}{2}\lp(R -\At_{ij} \At^{ij} 
        + \frac{2}{3} K^2 - 2\Theta K +2 D_i \zh^i 
        \rp.
        \\ \nonumber
        &\mathopen{}\lp. \hphantom{=}
        -2 \zh^i D_i\ln{\alpha} -16\pi\rho \rp),
\\
\label{rccz4_gh_fccz4}
        \lm \gt_{ij} &= -2 \alpha \At_{ij} - \frac{2}{3} \gt_{ij} \dt_{m}
        \beta^{m},
\\
\label{rccz4_ah_fccz4}
        \lm \At_{ij} &= - \frac{2}{3} \At_{ij} \dt_m \beta^m 
         +\alpha\At_{ij}\lp( K-2\Theta\rp)
        \\ \nonumber
        & \hphantom{=} 
        +\e4c \lp[-D_{i}D_{j} \alpha +\alpha\lp( R_{ij} + 2 D_{(i} \zh_{j)} 
        \vphantom{\frac{}{}}
        \rp.\rp.
        \\ \nonumber
        &\mathopen{}\lp.\mathopen{}\lp.\vphantom{\frac{}{}} \hphantom{=}
        -8\pi S_{ij} \rp)  \rp]^\tf 
        -2\alpha\At_{ik}\At^{k}\vp_{j},
\\
\label{rccz4_Lh_fccz4}
        \lm \Lamt^i &= \gt^{mn} \Do_m \Do_n \beta^i + \frac{2}{3} 
        \Lamt^i\dt_n\beta^n
        + \frac{1}{3}\dt^i\dt_n\beta^n 
        \\ \nonumber
        & \hphantom{=}
        - 2\At^{ik}\lp( \dt_k\alpha 
        - 6\alpha\dt_k\chi \rp) + 2\alpha\At^{jk}\Dt^{i}\vp_{jk} 
        \\ \nonumber
        & \hphantom{=}
        - \frac{4}{3} \alpha \dt^i K + 2\gt^{ik} \lp( \alpha\dt_k 
        \Theta - \Theta\dt_k \alpha 
        \rp.
        \\ \nonumber
        &\mathopen{}\lp. \hphantom{=}
        - \frac{2}{3}\alpha K \zh_k \rp)
        -16\pi\alpha\gt^{ij}j_j,
\\
\label{rccz4_Zh_fccz4}
        \lm \zh_i &= \alpha \lp( D_j {\At^{j}}_i -\frac{2}{3} D_i K +D_i\Theta 
        - \Theta D_i \ln\alpha
        \rp.
        \\ \nonumber
        &\mathopen{}\lp. \hphantom{=}
        -2 \zh_j {\At^{j}}_i - \frac{2}{3}\zh_i K -8 \pi j_i\rp),
\end{align}
where, as with RCCZ4, either $\zh_i$ or $\Lamt^i$ may be viewed as the 
fundamental dynamical quantity and the two are related via
\begin{align}
   \Lamt^i = \Dt^i + 2\gt^{ij} \zh_j.
\end{align}

The equations of motion for GBSSN, meanwhile, are:
\begin{align}
\label{rccz4_chi_bssn}
        \lm \chi &= -\frac{1}{6} \alpha K + \frac{1}{6}  \dt_m\beta^m,
\\
\label{rccz4_K_bssn}
        \lm K &= -D^2\alpha + \alpha\lp( \At_{ij} \At^{ij} 
        + \frac{1}{3} K^2 \rp) 
        \\ \nonumber 
        & \hphantom{=}
        + 4\pi \alpha\lp( \rho + S \rp),
\\
\label{rccz4_gh_bssn}
        \lm \gt_{ij} &= -2 \alpha \At_{ij} - \frac{2}{3} \gt_{ij} \dt_{m}
        \beta^{m},
\\
\label{rccz4_ah_bssn}
        \lm \At_{ij} &= \e4c \lp[-D_i D_j\alpha  + \alpha R_{ij} 
        - 8\pi\alpha S_{ij}\rp]^{\mathrm{TF}} 
        \\ \nonumber
        & \hphantom{=}
        - \frac{2}{3}\At_{ij}\dt_m\beta^{m} + \alpha\lp( K \At_{ij} - 
        2\At_{ik} \At^{k}\vp_{j} \rp),
\\
\label{rccz4_Dh_bssn}
        \lm \Lamt^i &= \gt^{mn} \Do_m \Do_n \beta^i -2\At^{im}\dt_m\alpha 
        \\ \nonumber
        & \hphantom{=}
        +2\alpha\lp( 6 \At^{ij} \dt_j \chi - \frac{2}{3} \gt^{ij}\dt_j 
        K - 8 \pi \jt^i \rp)
        \\ \nonumber
        & \hphantom{=}
        + \frac{1}{3} \lp[ \dt^i\lp(\dt_n \beta^n\rp) + 2\Lamt^i 
        \dt_n\beta^n\rp]
        \\ \nonumber
        & \hphantom{=}
        +2 \alpha \At^{mn} \Dt^{i}\vp_{mn}, 
\end{align}
where we note that we have replaced the usual variable $\Dt^i$ with 
$\Lamt^i$ for notational consistency when comparing to FCCZ4 and RCCZ4.
Note that in the evaluation of GBSSN dynamical quantities $\Lamt^i$ is
substituted for $\Dt^i$, such that~(\ref{rccz4_R_tilde_def}) becomes
\begin{align}
\label{rccz4_R_tilde_GBSSN_def}
   \Rt_{ij} &= -\frac{1}{2}\gt^{mn}\Do_m\Do_n\gt_{ij} + \gt_{m(i}\Do_{j)}
   \Lamt^m 
   \\ \nonumber
   & \hphantom{=}
   + \Lamt^m\Dt_{(ij)m} +2\Dt^{mn}\vp_{(i}\Dt_{j)mn} + \Dt^{mn}\vp_{i}
   \Dt_{mnj}.
\end{align}

\section{\label{rccz4_sec_comparison}Comparison of GBSSN, FCCZ4 and RCCZ4}

This section presents the results of three strong field tests that
compare RCCZ4 to FCCZ4 and GBSSN in spherical symmetry using a 
massless scalar field matter source.
In Sec.~\ref{rccz4_subsec_convergence} we investigate the convergence of each 
formalism for  subcritical initial data on uniform grids. 
Sec.~\ref{rccz4_subsec_black_hole} then studies the relative 
performance of each method in simulating black hole spacetimes with puncture 
initial data~\cite{hannam2008wormholes, alcubierre2008introduction}. Finally, 
Sec.~\ref{rccz4_subsec_critical_collapse} investigates the performance of 
each formalism in the context of critical collapse, where we tune to the 
threshold of black hole formation using adaptive mesh refinement (AMR).

For all investigations, we work in spherical symmetry with conformal spatial 
metric, $\gt_{ij}$,
\begin{align}
   \gt_{ij}&=
   \begin{bmatrix}
       g_a(t,r)   & 0            & 0    \\
       0          & r^2 g_b(t,r) & 0    \\
       0          & 0            & r^2 \sin^2\theta g_b(t,r)
   \end{bmatrix},
\end{align}
unit normal, $n^{\mu}$, to the foliation,
\begin{align}
   n^\mu &= \frac{1}{\alpha(t,r)}
   \begin{bmatrix}
       1  & -r \beta_a(t,r) & 0 & 0
   \end{bmatrix},
\end{align}
trace-free extrinsic curvature, ${\At^{i}}_{j}$,
\begin{align}
   {\At^{i}}_{j}&=
   \begin{bmatrix}
       A_a(t,r)   & 0            & 0    \\
       0          & A_b(t,r)     & 0    \\
       0          & 0            & A_b(t,r)
   \end{bmatrix},
\end{align}
stress tensor, ${S^{i}}_j$,
\begin{align}
   {S^{i}}_{j}&=
   \begin{bmatrix}
       S_a(t,r)   & 0            & 0    \\
       0          & S_b(t,r)     & 0    \\
       0          & 0            & S_b(t,r)
   \end{bmatrix},
\end{align}
momentum density, $j^{i}$,
\begin{align}
   j^i &= 
   \begin{bmatrix}
      rj_a(t,r) & 0 & 0
   \end{bmatrix},
\end{align}
conformal connection functions $\Dt^{i}$ and $\Lamt^i$,
\begin{align}
   \Dt^i &= 
   \begin{bmatrix}
      r\Dt_a(t,r) & 0 & 0
   \end{bmatrix},
\\
   \Lamt^i &= 
   \begin{bmatrix}
      r\Lamt_a(t,r) & 0 & 0
   \end{bmatrix},
\end{align}
and spatial projections of $Z_\mu$,
\begin{align}
   \zh_i &= 
   \begin{bmatrix}
      r\zh_a(t,r) & 0 & 0
   \end{bmatrix}.
\end{align}
We take a massless scalar field, $\psi(t,r)$, with stress-energy tensor,
\begin{align}
   T_{\mu \nu} =\n_\mu\psi \n_\nu \psi -\frac{1}{2}g_{\mu\nu} \n_\sigma\psi \n^\sigma\psi,
\end{align}
as our matter model. 

The equations of motion are found through application of the results of
Secs.~\ref{rccz4_sec_derivation} and \ref{rccz4_sec_GBSSN_FCCZ4_EoM}. 
In order to regularize the equation of motion in the vicinity of black hole 
punctures, we evolve the regular quantity $X=e^{-2\chi}$ in place of $\chi$. 
As defined above, all of $\alpha$, $\beta_a$, $g_a$, $g_b$, $X$, $\At_a$, 
$\At_b$, $K$, $\Dt_a$, $\Lamt_a$, $\Theta$, $\zh_a$, $\rho$, $\jt_a$, $S$, 
$S_a$ and $S_b$ are even functions of $r$ as $r\rightarrow0$ and the 
following identities hold:
\begin{align}
   g_a &= \frac{1}{g_b^2},
\\
   A_a &= -2 A_b.
\end{align}

\subsection{Convergence and Independent Residual Tests
\label{rccz4_subsec_convergence}}

We validate our evolution schemes and code through the use of independent 
residual convergence and by monitoring the convergence of various 
constraints. 
All tests are performed for marginally subcritical initial data so that 
slightly stronger initial data would result in black hole formation.

Our code is implemented as a simple second order in space and time 
Crank-Nicolson solver using a uniform grid in $r$ and $t$ with fourth order 
Kreiss-Oliger dissipation~\cite{kreiss1973methods} applied at the current and 
advanced time levels. The code is built 
on {PAMR}~\cite{pamr_reference} and {AMRD}~\cite{amrd_reference} and supports 
AMR in space and time using the Berger-Oliger 
approach~\cite{berger1984adaptive}. Grid function values at refinement 
boundaries are set via third order temporal interpolation. 

Our independent residual evaluators take the form of alternative 
discretizations of the ADM equations applied to our computed solutions. 
The application of these alternative discretizations helps to ensure that our 
evolution scheme is free of subtle flaws while our use of the ADM equations 
(as opposed to GBSSN, FCCZ4 or RCCZ4), aids in demonstrating convergence to 
GR rather than some other differential system. 

Returning to the specific calculations performed in this subsection, 
the initial data is taken to be time 
symmetric with the massless scalar field, $\psi$, set according to:
\begin{align}
   \psi(0,r) &= a e^{-\lp( r - r_0 \rp)^2/\sigma^2},
\\
   \p_t\psi(0,r)  &= 0.
\end{align}
Specifically, for our testing we have taken $a = 0.035$, $\sigma = 2$ and 
$r_0 = 12$ so that, as mentioned above, we are in the subcritical regime but 
relatively close to the critical point of $a \approx 0.0362$. The dynamics 
are therefore non-linear, span several orders of magnitude, and are far 
from trivial.
Initial data for the conformal factor, $X=e^{-2\chi}$, is determined by 
solving the Hamiltonian constraint on a finite grid where $X$ is assumed to 
behave as $1+a/r$ at the outer boundary. This grid is sized so that errors 
at the outer boundary are unable to propagate into the region of interest 
during the course of the convergence testing.

Our simulations are run with generalized 1+log lapses and a Lambda driver 
shift given by
\begin{align}
\label{rccz4_1plog_bssn}
   \p_t \alpha &= -2\alpha K,
\\ 
\label{rccz4_1plog_ccz4}
   \p_t \alpha &= -2 \alpha \lp( K - 2\Theta \rp) ,
\\ 
\label{rccz4_1plog_rccz4}
   \p_t \alpha &= -2 \alpha \lp( K - 2\Thetat \rp),
\\ 
\label{rccz4_delta_driver}
   \p_{tt} \beta^i &= \frac{3}{4}\p_t \Lambda^i -2\p_t\beta^i ,
\end{align}
where (\ref{rccz4_1plog_bssn}), (\ref{rccz4_1plog_ccz4}) and 
(\ref{rccz4_1plog_rccz4}) are the slicing conditions used for GBSSN, 
FCCZ4 and RCCZ4, respectively.

\begin{figure}[!ht]
\centering
\includegraphics[scale=1.0]
{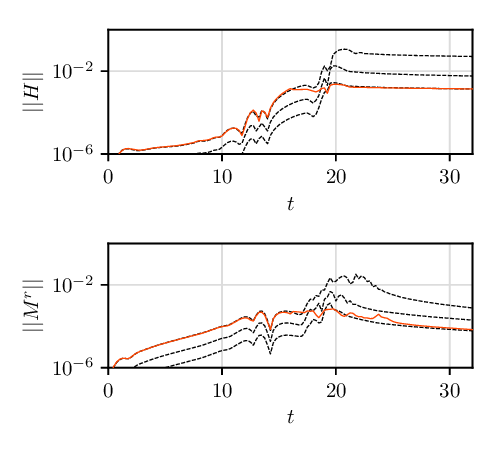}
\caption{$l_2$ norms of the Hamiltonian and momentum constraint violations for
the GBSSN formulation. Simulations are shown for fixed resolutions (dashed
lines) of 1025, 2049 and 4097 points. Results from an AMR simulation with a 
relative local error tolerance of  $10^{-4}$ are shown as the solid colored 
lines. The AMR simulations are well within the convergent regime. }
\label{rccz4_methods_bssn_converge}
\end{figure}

\begin{figure}[!ht]
\centering
\includegraphics[scale=1.0]
{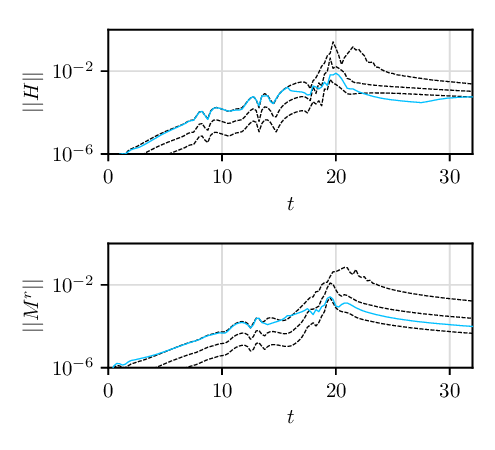}
\caption{$l_2$ norms of the Hamiltonian and momentum constraint violations 
for the FCCZ4 formulation. Simulations are shown for fixed resolutions 
(dashed lines) of 1025, 2049 and 4097 points. Results from an AMR simulation 
with a relative local error tolerance of  $10^{-4}$ are shown as the solid 
colored lines. The AMR simulations are observed to be well within the 
convergent regime. }
\label{rccz4_methods_ccz4_converge}
\end{figure}

\begin{figure}[!ht]
\centering
\includegraphics[scale=1.0]
{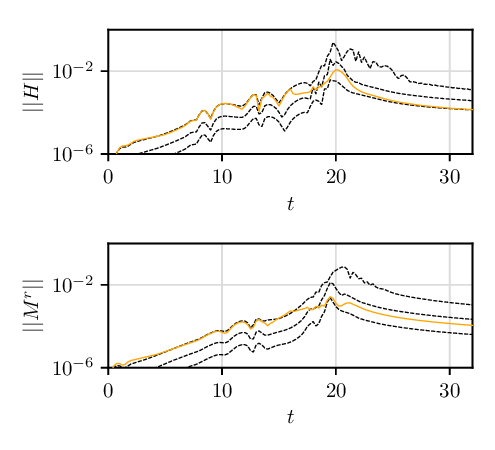}
\caption{$l_2$ norms of the Hamiltonian and momentum constraint violations 
for the RCCZ4 formulation. Simulations are shown for fixed resolutions 
(dashed lines) of 1025, 2049 and 4097 points. Results from an AMR simulation 
with a relative local error tolerance of  $10^{-4}$ are shown as the solid 
colored lines. The AMR simulations are well within the convergent regime. }
\label{rccz4_methods_rccz4_converge}
\end{figure}

Figures~\ref{rccz4_methods_bssn_converge}--\ref{rccz4_methods_rccz4_converge} 
demonstrate convergence of the Hamiltonian and momentum constraints for each 
of GBSSN, FCCZ4 and RCCZ4. In each figure, the dashed lines show norms 
evaluated on a $r=[0, 64]$ grid at fixed resolutions of 1025, 2049 and 4097 
points, respectively. AMR calculations with a per-step error tolerance of 
$10^{-4}$ are shown with solid lines and demonstrate that with an appropriate 
choice of parameters, the adaptive computations remain within the convergent 
regime.
For each simulation, and prior to the evaluation of their norm, the 
constraints are interpolated to a uniform grid of fixed resolution. This 
enables direct comparison of the convergence rates among the simulations. In 
these figures, a factor of 4 difference in the independent residuals or 
constraint maintenance between runs which differ by a factor of 2 in grid 
spacing indicates second order convergence.

\begin{figure}[!ht]
\centering
\includegraphics[scale=1.0]
{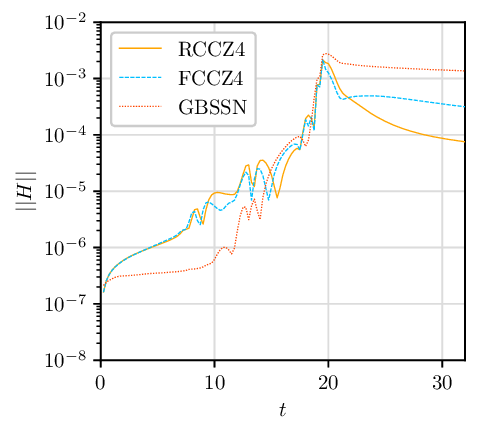}
\caption{$l_2$ norm of the Hamiltonian constraint violation for the case of
strong field initial data for each of GBSSN, FCCZ4 and RCCZ4. The difference
between RCCZ4 and FCCZ4 is largely due to a more pronounced outgoing pulse of
constraint violation (which leaves nearly flat space in its wake) while the
large static constraint violation of GBSSN is concentrated at the origin and
leaves behind a metric that does not appear to be a valid solution to the 
Einstein-scalar equations. }
\label{rccz4_methods_comp_ham}
\end{figure}

\begin{figure}[!ht]
\centering
\includegraphics[scale=1.0]
{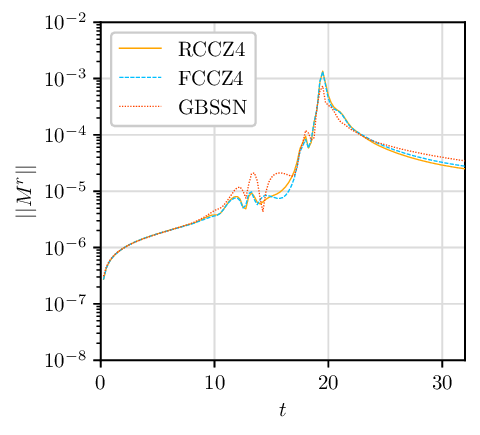}
\caption{$l_2$ norm of the momentum constraint violation for the case of strong
field initial data for each of GBSSN, FCCZ4 and RCCZ4. Not surprisingly, the
performance of the three methods is largely equivalent as they 
are all designed to advect away
the momentum constraint violation. }
\label{rccz4_methods_comp_mom}
\end{figure}

\begin{figure}[!ht]
\centering
\includegraphics[scale=1.0]
{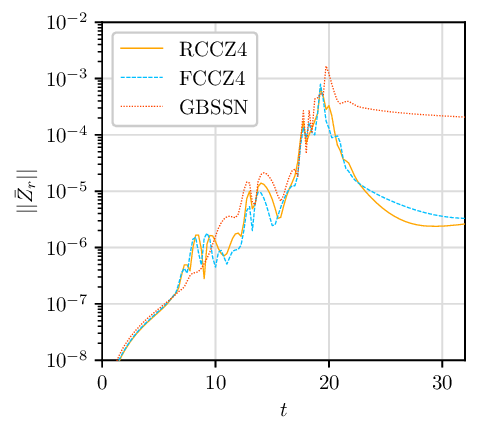}
\caption{$l_2$ norm of $\zh_r = g_a( \protect\Lamt^r
-\protect\Dt^r)/2$  for the case of strong field initial data for each of 
GBSSN, FCCZ4 and RCCZ4. As in the case of the Hamiltonian constraint, the GBSSN 
errors are concentrated at the origin where the curvature takes on its 
largest values. This error remains essentially static save for the mitigating 
factor of dissipation. At this resolution, FCCZ4 preserves the constraint 
about 100 times better than GBSSN while RCCZ4 improves upon this by a further 
factor of $\sim3$ or so at late times.
}
\label{rccz4_methods_comp_z4_clipped}
\end{figure}

\begin{figure}[!ht]
\centering
\includegraphics[scale=1.0]
{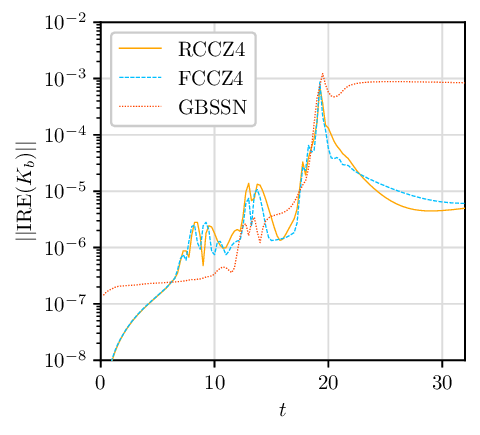}
\caption{$l_2$ norm of the independent residual evaluator for $K_b$. At late
times, as the solution should be approaching flat space, RCCZ4 has better
performance than either FCCZ4 or GBSSN. }
\label{rccz4_methods_comp_kb_clipped}
\end{figure}

Figs.~\ref{rccz4_methods_comp_ham}--\ref{rccz4_methods_comp_kb_clipped} show
the performance of each formalism relative to one another. The simulations are
run at a resolution of 4097 grid points on a grid which extends to $r = 64$ 
(corresponding to the most refined unigrid run of
Figs.~\ref{rccz4_methods_bssn_converge}--\ref{rccz4_methods_rccz4_converge}).
We choose the domain on which the norms are evaluated such that signals 
have not had sufficient time to propagate from the outer boundary (which is 
set assuming $X=1+a/r$ for some value $a$) into the domain of interest.
It should be stressed that for all of the norms plotted in
Figs.~\ref{rccz4_methods_comp_ham}--\ref{rccz4_methods_comp_kb_clipped}, the
solutions are well resolved. 
The significant, and previously studied, improvements of the FCCZ4 method 
over GBSSN~\cite{daverio2018apples} in maintaining the Hamiltonian 
constraint and independent residuals is a real effect which is present even 
at high resolutions.

\subsection{Evolution of Black Hole Spacetimes
\label{rccz4_subsec_black_hole}}

In order for RCCZ4 (or a to-be-developed formalism based upon similar
principles) to be competitive with GBSSN or FCCZ4 in the domain of strong 
field numerical simulations (which frequently involve singularities), it 
first needs to be capable of stably evolving black holes.
Here, we show that with minor modifications to the standard 1+log and Delta 
driver gauges, RCCZ4 in spherical symmetry is at least as capable as FCCZ4 
for the evolution of black hole space times.

We start with standard time symmetric, black hole puncture initial 
data~\cite{alcubierre2008introduction, gourgoulhon2012} given by:
\begin{align}
   X &= \lp(1 + \frac{M}{2r} \rp)^{-2} ,
\\
   \alpha &= \lp(1 + \frac{M}{2r} \rp)^{-2} ,
\\
   \beta_a &= K = A_a = A_b = 0,
\\
   g_a &= g_b = 1.
\end{align}

The simulations are performed on large grids ($r = [0, 128 M]$ with $M = 4$) 
which are further refined via fixed mesh refinement (FMR)~\footnote{For these 
tests we wanted to have as little contamination from imperfectly specified 
boundary conditions as possible while performing long term evolutions. 
Correspondingly, we placed the outer boundary at $r=128M$ and evolved until 
$t = 64M$. All results presented 
are evaluated on the portion of the spatial domain between the horizon and 
$r = 8M$. }.
The sizes of the fixed refinement regions were determined by first 
evolving the initial data with adaptive mesh refinement. At the 
conclusion of this AMR run, each level of refinement had a maximum
extent and that maximum extent then defined the limits of the corresponding 
level of refinement for the FMR calculations. 
In the simulations, the use of mesh refinement serves several purposes. 
First, it reduces the computational load for high resolution simulations. 
Second, it allows us to verify the compatibility of our implementation of 
the {GBSSN}, {FCCZ4} and {RCCZ4} formalisms with AMR. Third, by using 
fixed (as opposed to adaptive) mesh refinement, we eliminate complications 
caused by each formulation employing slightly different regridding 
procedures. This, in turn, facilitates the analysis of convergence properties. 
Table~\ref{rccz4_bh_meshes} shows the extent and refinement ratio of each 
grid used for the black hole simulations.

We note that the quantities $\zh_i$ and $\Theta$ are effectively error terms 
which serve to propagate violations of the momentum and Hamiltonian constraints 
and that they tend to 
grow in the vicinity of refinement boundaries. As such, we find that is is 
best to either evolve $\Lamt^i$ (rather than $\zh_i$) or to omit $\Theta$ and 
$\zh_i$ from the truncation error calculation used to determine the placement 
of refined regions.

\begin{table}[!ht]
\centering
\begin{ruledtabular}
\begin{tabular}{ c  c  c  c  c  c  c }
    Level & $r_{\rm min}$ & $r_{\rm max}$ & $h^0_{r}$ & $h^1_{r}$ & $h^2_{r}$ 
    & $h^3_{r}$\\
    \hline
    1     & 0 & 512 & 8         & 4        & 2        & 1         \\  
    2     & 0 & 512 & 4         & 2        & 1        & $2^{-1}$  \\
    3     & 0 & 256 & 2         & 1        & $2^{-1}$ & $2^{-2}$  \\
    4     & 0 & 256 & 1         & $2^{-1}$ & $2^{-2}$ & $2^{-3}$  \\
    5     & 0 & 128 & $2^{-1}$  & $2^{-2}$ & $2^{-3}$ & $2^{-4}$  \\
    6     & 0 & 64  & $2^{-2}$  & $2^{-3}$ & $2^{-4}$ & $2^{-5}$  \\
    7     & 0 & 32  & $2^{-3}$  & $2^{-4}$ & $2^{-5}$ & $2^{-6}$  \\    
\end{tabular}
\end{ruledtabular}
\caption{ 
Parameters for the meshes in fixed mesh refinement convergence simulations. 
The fixed mesh refinement simulations use a total of 7 refinement 
levels as labeled in the first column. The extent of each mesh is displayed 
in columns 2 and 3 ($r_{\rm min}$ and $r_{\rm max}$). The grid spacings for the 
lowest resolution simulation are shown in the fourth column ($h^0_r$). Each 
of the final three columns ($h^1_r$, $h^2_r$ and $h^3_r$) 
give grid spacings for 
progressively higher resolution simulations. As an example, the 
$6^{\mathrm{th}}$ refinement level (Level 6) has a spatial extent of 
$r=[0, 64]$. For the most resolved simulation ($h^3_r$), the grid spacing 
on that level is $2^{-5}$. 
 }
\label{rccz4_bh_meshes}
\end{table}

\begin{figure}[!ht]
\centering
\includegraphics[scale=1.0]
{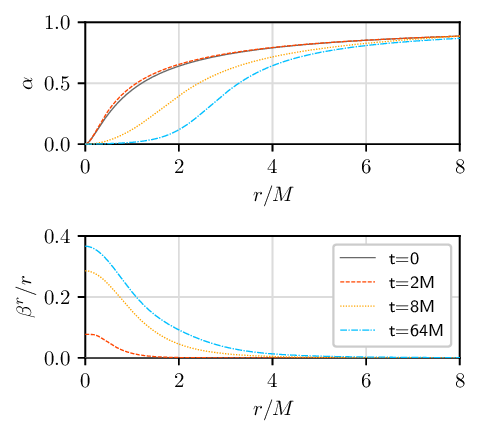}
\caption{ Evolution of $\alpha$ and $\beta^r$ 
from $t=0$ to $t=64M=256$. The initial puncture type initial data quickly
evolves towards trumpet type data with $\alpha$ going as 
$r$ as opposed to $r^2$ at the puncture.}
\label{rccz4_bh_time_1}
\end{figure}

\begin{figure}[!ht]
\centering
\includegraphics[scale=1.0]
{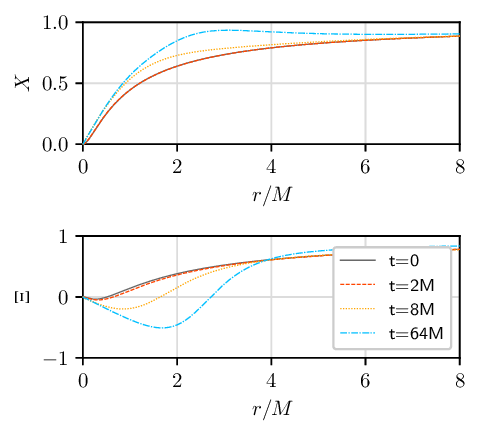}
\caption{ Evolution of $X$ and $\Xi$ 
from $t=0$ to $t=64M=256$. The initial puncture type initial data quickly
evolves towards trumpet type initial data with $X$ going as $r$ as opposed 
to $r^2$ at the puncture. As can be seen in the graph of 
$\Xi$, the coordinate location of the apparent horizon 
(where $\Xi = 0$) 
increases slowly with coordinate time.
}
\label{rccz4_bh_time_2}
\end{figure}

Figures~\ref{rccz4_bh_time_1}--\ref{rccz4_bh_time_2}  
show the evolutions of $\alpha$, $\beta^r$ and
$X$ as well as the coordinate location of the apparent horizon.
\textcolor{black}{determined by a zero of the quantity $\Xi$:}
\begin{align}
   \Xi &= \frac{r X \p_r g_b}{2} - rg_b\p_r X + g_b X - \frac{r\lp(K+3A_b\rp)}{3}.
\end{align} 
As is well known \cite{alcubierre2003_mp_5,
hannam2008wormholes, brown2008bssn, hilditch2013collapse,
thierfelder2011trumpet, alcubierre2011formulations}, puncture type initial
data evolves towards a trumpet like spacetime and performs a form of automatic
excision in the vicinity of the puncture. In this region, the  evolved and 
constrained quantities do not converge. 

\begin{figure}[!ht]
\centering
\includegraphics[scale=1.0]
{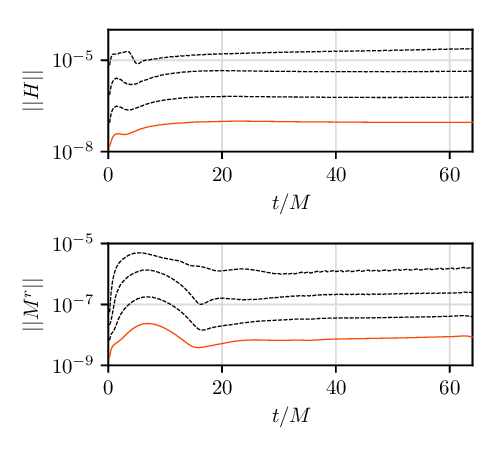}
\caption{$l_2$ norms of the Hamiltonian and momentum constraint violations for
the GBSSN formulation. Each successive line denotes a factor of 2 grid
refinement. The solid line denotes the most refined simulation.}
\label{rccz4_bh_bssn_converge}
\end{figure}

\begin{figure}[!ht]
\centering
\includegraphics[scale=1.0]
{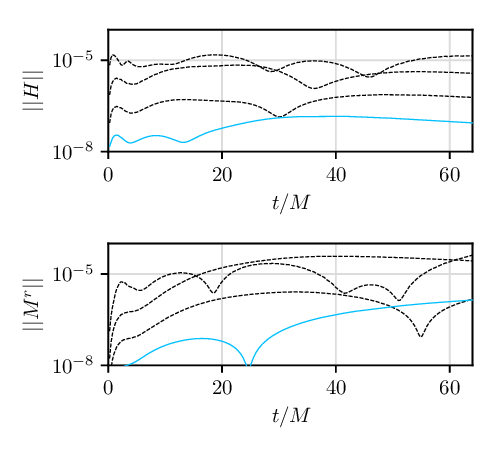}
\caption{$l_2$ norms of the Hamiltonian and momentum constraint violations 
for the FCCZ4 formulation. The errors in the momentum constraint appear 
to be dominated by artifacts that arise at the mesh refinement boundaries. 
Our GBSSN and RCCZ4 simulations used identical parameters and neither 
experienced the same sort of issues arising at the mesh refinement 
boundaries. Rather than attempting to find more optimal parameters which 
could resolve these issues at the cost of preventing direct comparison with 
GBSSN and RCCZ4, the simulation is left as-is and we note that it would 
almost certainly be possible to find better parameters for FCCZ4
which would mitigate these issues. }
\label{rccz4_bh_ccz4_converge}
\end{figure}

\begin{figure}[!ht]
\centering
\includegraphics[scale=1.0]
{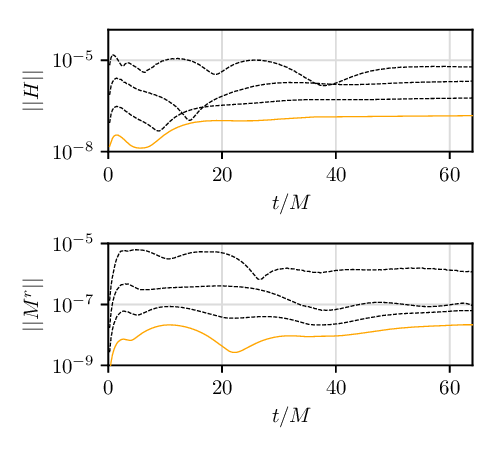}
\caption{$l_2$ norms of the Hamiltonian and momentum constraint violations for
the RCCZ4 formulation. Each successive line denotes a factor of 2 grid
refinement. The solid line denotes the most refined simulation. }
\label{rccz4_bh_rccz4_converge}
\end{figure}

\begin{figure}[!ht]
\centering
\includegraphics[scale=1.0]
{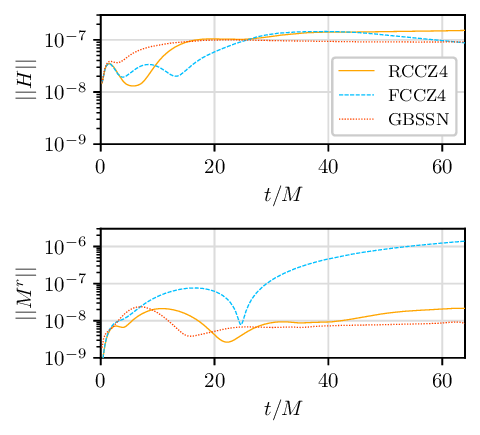}
\caption{$l_2$ norms of the Hamiltonian constraint and momentum constraint 
violation for the $h_r = h^3_r$ run of each of the RCCZ4, FCCZ4 and GBSSN 
formulations.  Here we can observe key differences in the constraint 
violating behaviours of each formulation. As the GBSSN simulation does not 
couple the Hamiltonian constraint to a propagating degree of freedom, errors 
within the horizon and at refinement boundaries are unable to propagate. Due 
to the fact that the black hole is not moving and the simulation quickly 
approaches a nearly stationary state, this lack of time dependence is 
advantageous. As shown in Section~\ref{rccz4_subsec_convergence}, the 
opposite is true when the simulation is highly dynamic. In those cases, both 
RCCZ4 and FCCZ4 provide orders of magnitude better constraint conservation. }
\label{rccz4_bh_rccz4_ccz4_bssn}
\end{figure}

The convergence of the $l_2$ norms of the various constraints in the region
external to the apparent horizon ($r = [r_{\mathrm{AH}}, 8M]$) and for each
formalism are shown in 
Figs.~\ref{rccz4_bh_bssn_converge}--\ref{rccz4_bh_rccz4_converge}. 
The dashed lines show simulations with $h_r = h^0_r$, $h_r = h^1_r$ and 
$h_r = h^2_r$ while the solid color denotes the most resolved $h_r = h^3_r$ 
simulation. Fig.~\ref{rccz4_bh_rccz4_ccz4_bssn} compiles the highest 
resolution runs of 
Figs.~\ref{rccz4_bh_bssn_converge}--\ref{rccz4_bh_rccz4_converge} and permits 
a direct comparison of the implementations. Independent residuals
behave similarly and so have not been plotted.

Examining Fig.~\ref{rccz4_bh_rccz4_ccz4_bssn}, we see that for a stationary 
black hole, GBSSN is favoured over either FCCZ4 or RCCZ4. For a simulation 
where we are concerned with computing the constraint violation external to 
the apparent horizon, this makes intuitive sense: the formulation which does 
not propagate Hamiltonian constraint violations away from punctures or grid 
refinement boundaries should produce superior results when the fields are 
nearly stationary. However, as shown in~\cite{hilditch2013compact}, for more 
dynamical situations we should not expect superior performance from BSSN-type
\textcolor{black}{
simulations even when constraint damping is employed.}

As noted in Fig.~\ref{rccz4_bh_ccz4_converge}, the errors in 
the momentum constraint (and $\zh_r$) for FCCZ4 appear to be dominated by
the development of artifacts at the mesh refinement boundaries. Doubtless, 
these issues could be mitigated with proper attention. The relatively poor 
performance of FCCZ4 in comparison to GBSSN and RCCZ4 in these simulations 
should therefore not be seen as a shortcoming of the method, but as an issue 
arising from our demand that the methods be compared via runs with identical 
parameters. Taking this into account, we see that at early times (before the 
errors become dominated by issues arising from grid refinement boundaries), 
the performance of each method is roughly equivalent.

\subsection{\label{rccz4_subsec_critical_collapse}Critical Collapse}

Critical collapse represents the extreme strong field regime of general 
relativity and is therefore an excellent test case to determine the 
capabilities of a numerical formulation. Here we compare the {RCCZ4}, 
{FCCZ4} and {GBSSN} formalisms, without constraint damping, in a test that 
studies each formalism's capacity to resolve the threshold of black hole 
formation using gauges which are natural extensions of the 1+log slicing, 
(\ref{rccz4_1plog_bssn}--\ref{rccz4_1plog_rccz4}), with zero shift. For 
additional information concerning critical collapse, 
see~\cite{choptuik_universality_and_scaling_1993} for the original study 
concerning the massless scalar field in spherical symmetry 
and~\cite{carsten1999critical, gundlach_critical_phenomena_2007} for more 
general reviews.

For each of GBSSN, FCCZ4 and RCCZ4, we perform AMR simulations of massless 
scalar field collapse with a relative, per-step truncation error tolerance 
of $10^{-4}$. We tune the amplitude of our initial data to the threshold of 
black hole formation with a relative tolerance of $\sim10^{-12}$. 

\begin{figure}[!ht]
\centering
\includegraphics[scale=1.0]
{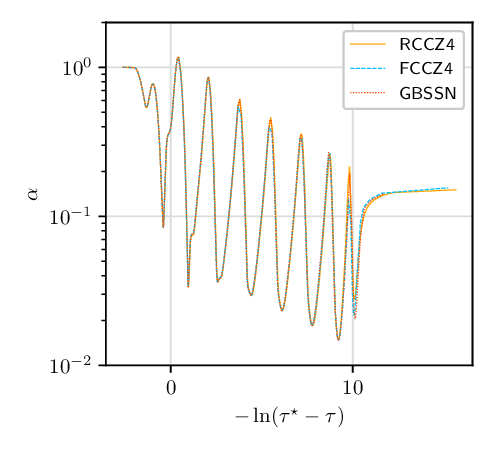}
\caption{Lapse, $\alpha$, at the accumulation point as a function of 
$-\ln(\tau^\star-\tau)$ with $\tau^\star$ an approximate accumulation time 
which is different for each set of simulations. Each of GBSSN, FCCZ4 and RCCZ4 
are well suited to performing the critical evolutions. The observed 
discrepancies in $\alpha$ are primarily due to our output of data with 
insufficient frequency to resolve the peaks adequately. As expected, 
we are able to resolve approximately 3 echos at a relative search tolerance 
of $10^{-12}$. }
\label{rccz4_critical_alpha}
\end{figure}

\begin{figure}[!ht]
\centering
\includegraphics[scale=1.0]
{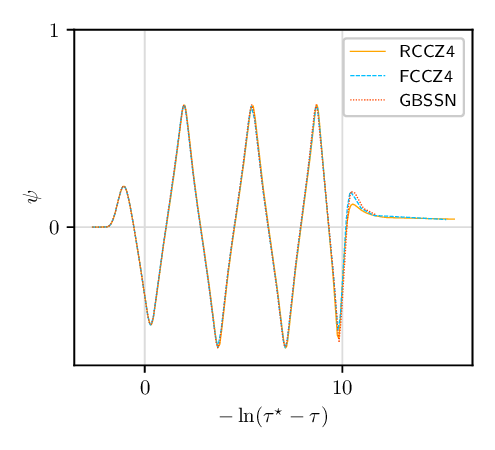}
\caption{Scalar field, $\psi$, at the accumulation point as a function of 
$-\ln(\tau^\star-\tau)$. The discrete self similarity (DSS) is evident. Tuning 
the amplitude of our initial data to the threshold of black hole formation 
with a relative tolerance of $\sim10^{-12}$ allows us to resolve 
approximately three echos.
}
\label{rccz4_critical_psi}
\end{figure}

\begin{figure}[!ht]
\centering
\includegraphics[scale=1.0]
{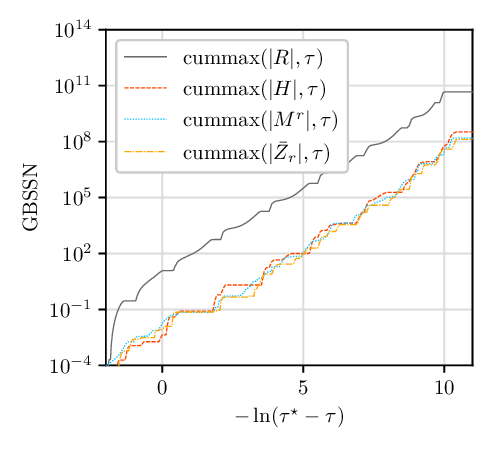}
\caption{Cumulative maximal values of $R$, $\zh_r$, the Hamiltonian constraint and
momentum constraint violations for critical collapse of the scalar field in
the GBSSN formulation. For clarity, we have not shown the behaviour of the 
Hamiltonian constraint post-dispersal, where it is dominated by a large non 
propagating remnant similar to that seen in 
Fig.~\ref{rccz4_methods_bssn_converge}. }
\label{rccz4_critical_bssn}
\end{figure}

\begin{figure}[!ht]
\centering
\includegraphics[scale=1.0]
{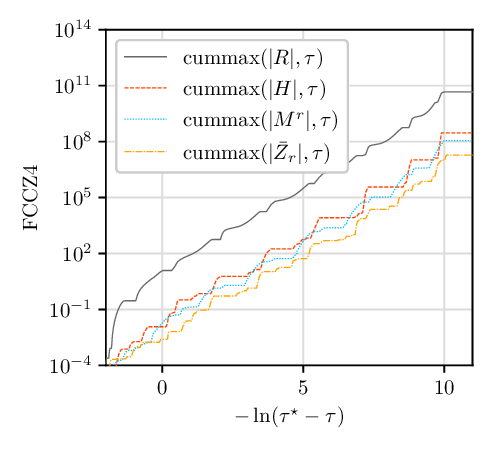}
\caption{ Cumulative maximal values of $R$, $\zh_r$, the Hamiltonian constraint
and momentum constraint violations for critical collapse of the scalar field
in the FCCZ4 formulation. For subcritical simulations close to criticality,
the post dispersal constraint violating remnant is much smaller than that of
GBSSN but is still too large to continue the simulation for long periods of 
time. }
\label{rccz4_critical_ccz4}
\end{figure}

\begin{figure}[!ht]
\centering
\includegraphics[scale=1.0]
{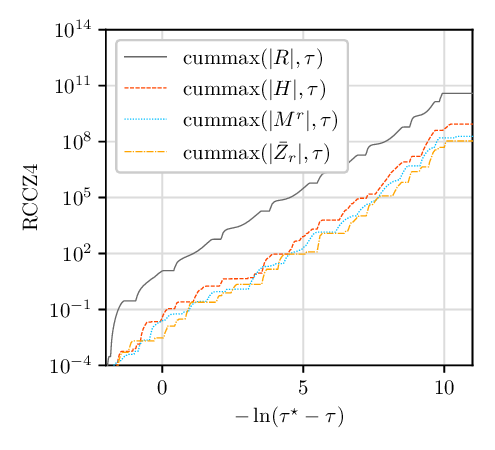}
\caption{ Cumulative maximal values of $R$, $\zh_r$, the Hamiltonian constraint
and momentum constraint violations for critical collapse of the scalar
field in the RCCZ4 formulation. For subcritical simulations close to
criticality, the post dispersal constraint violating remnant is much
smaller than that of GBSSN but is still too large to continue the simulation 
for long periods of time. Close to criticality, the constraint
violations grow noticeably faster than either GBSSN or FCCZ4 (while still
providing adequate resolution to investigate criticality). 
}
\label{rccz4_critical_rccz4}
\end{figure}

Figures~\ref{rccz4_critical_alpha}--\ref{rccz4_critical_psi} plot the central 
value of the lapse and the scalar field, respectively, 
against proper time at the approximate 
accumulation point (the spacetime point at which a naked singularity would 
form in the limit of infinite tuning) for 
the subcritical simulation closest to criticality in each formalism. 
Figures.~\ref{rccz4_critical_bssn}--\ref{rccz4_critical_rccz4} plot the 
magnitudes of constraint violations from these calculations.
For these simulations we expect all dimensionful quantities to 
grow exponentially in $-\ln{(\tau^{\star} -\tau)}$ due to the discretely 
self-similar nature of the critical solution. To facilitate analysis of the 
overall growth rate of constraint violations, we plot the cumulative 
maximum, $\mathrm{cummax}(f(t), t)$, of each quantity. This function returns 
the largest magnitude encountered on the domain of the simulation up until 
that point in time (e.g. $\mathrm{cummax}(R, t_0)$ would return the largest 
value of $R$ encountered during the simulation for $t=[0,t_0]$).

As seen in Figs.~\ref{rccz4_critical_bssn}--\ref{rccz4_critical_rccz4}, when
evolved using identical error tolerances and parameters, we find that GBSSN
does the best at maintaining a constant level of relative constraint violation 
throughout the simulation. 
We find that with a per-step error tolerance of $10^{-4}$, GBSSN maintains a 
constant error ratio of about $10^{-3}$ relative to the magnitude of the 
Ricci scalar.  For FCCZ4, this is reduced to $10^{-2}$ while RCCZ4 performs 
similarly to FCCZ4 for the first echo or so and then gradually accumulates 
more error, performing worse than either GBSSN or FCCZ4 at late times. 

At this point, the cause of this dip in performance for RCCZ4 is unclear to 
us. However, it is entirely possible that it is due to a suboptimal 
regridding strategy. Alternatively, it could very well be that the variant of 
the 1+log slicing condition used, Eqn.~(\ref{rccz4_1plog_rccz4}), is 
not ideal for controlling the Hamiltonian constraint. 
We tried several variations of the form $\p_t \alpha = -2 \alpha ( K - 
2f(\alpha)\Thetat )$, which, for the most part, resulted in similar 
performance and stability properties. 

\textcolor{black}{
The superior performance of GBSSN in the approach to criticality contrasts 
with its poor performance post dispersal. As in 
Sec.~\ref{rccz4_subsec_convergence}, after a simulation achieves its closest 
approach to criticality, the scalar field disperses to infinity and would 
ideally leave flat space in its wake. Both FCCZ4 and RCCZ4 perform better 
than GBSSN in this regime although this is not evident when plotting 
cumulative maxima as in 
Figs.~\ref{rccz4_critical_bssn}--\ref{rccz4_critical_rccz4}.}

\FloatBarrier
\section{\label{rccz4_sec_hyperbolicity}Hyperbolicity of RCCZ4}

We now turn to an analysis of the hyperbolicity of RCCZ4. We demonstrate 
that, relative to GBSSN, RCCZ4 has one fewer zero-velocity modes, which 
roughly corresponds to the fact that in Z4 derived formulations the 
equivalent of the Hamiltonian constraint is 
dynamical~\cite{hilditch2013compact, daverio2018apples, 
bernuzzi2010constraint}. As outlined in~\cite{cao2022note, 
alcubierre2000towards, mongwane2016hyperbolicity}, and in the context of 
numerical relativity, these zero-velocity modes often correspond to 
constraint violations and are thought to contribute to instabilities. 
Consequently, formulations that minimize these modes are generally favoured.

Here we derive the conditions under which  RCCZ4 is hyperbolic, performing a 
pseudodifferential reduction~\cite{gundlach2006hyperbolicity, 
nagy2004strongly} following the procedure of Cao and Wu~\cite{cao2022note} 
who have previously applied the method to a study of the hyperbolicity of BSSN 
in $f(R)$ gravity. We consider the RCCZ4 equations of motion 
(\ref{rccz4_chi_rccz4})--(\ref{rccz4_Zh_rccz4}) in the vacuum
and choose a generalization of the Bona-Masso family of 
lapses~\cite{bona1995new, alcubierre2008introduction} together with 
generalized Lambda drivers for the shift.  Specifically, defining
\begin{align}
\p_0 &= \p_t - \beta^i\p_i
\end{align}
the equation for the lapse is 
\begin{align}
     \p_0 \alpha &= -\alpha^2 h\lp(\alpha, \chi\rp)\lp(\frac{}{}
     K - K_0 
     -\frac{m\lp(\alpha, \chi\rp)}{\alpha}\Thetat\rp) .
\end{align}
Our generalized Lambda driver takes the form
\begin{align}
     \p_0 \beta^i &= \alpha^2 G\lp(\alpha, \chi\rp)B^i,
\end{align}
where the auxiliary vector $B^i$ satisfies
\begin{align}
     \p_0 B^i &=e^{-4\chi} H\lp(\alpha, \chi\rp)\p_0\Lamt^i 
     - \eta\lp(B^i,\alpha\rp),
\end{align}
and $G$ and $H$ are some specified functions.

We wish to determine the conditions under which the RCCZ4 system is strongly 
hyperbolic. This essentially amounts to verifying that the system admits a 
well defined Cauchy problem; i.e.~that there exist no high frequency modes 
with growth rates which cannot be bounded by some exponential function of 
time~\cite{nagy2004strongly}. We can thus study strong hyperbolicity by 
linearizing the equations about some generic solution and examining the 
resulting
\textcolor{black}{
perturbed system in the high frequency regime where it takes the form
\begin{align}
   \label{rccz4_hyperlinearized}
   \p_0 u  = \mathbf{M}^i \p_i u  + \mathbf{S}u .
\end{align}
Here, $u$ is a vector of $n$ 
perturbation fields, }
$\mathbf{M}^i$ are $n$-by-$n$ 
characteristic matrices and $\mathbf{S}u$ is a source vector that may depend 
on the fundamental variables $u$ but not on their derivatives.
Fourier transforming the 
\textcolor{black}{
perturbation 
$u$ via}
\begin{align}
   \uh\lp(\omega\rp) = \int{e^{i(\omega_k x^k)} u\lp(x\rp)\mathrm{d}^3 x},
\end{align}
we can write (\ref{rccz4_hyperlinearized}) as
\begin{align}
   \p_0 \uh  = i\omega_i\mathbf{M}^i \uh  + \mathbf{S}\uh.
\end{align}
From this, we define the principal symbol of the system as $\mathbf{P}_1 = 
i\wabs\mathbf{P} = i\omega_i\mathbf{M}^i$. The hyperbolicity of the system 
can then be discerned from the properties of $\mathbf{P}$:
\begin{itemize}
   \item If $\mathbf{P}$ has imaginary eigenvalues, the system is not 
   hyperbolic and cannot be formulated as a well-posed Cauchy problem.
   \item If $\mathbf{P}$ has only real eigenvalues but does not possess a 
   complete set of eigenvectors, the system is weakly hyperbolic and may have 
   issues with ill-posedness.
   \item If $\mathbf{P}$ has both real eigenvalues and a complete set of 
   eigenvectors, the system is strongly hyperbolic and the Cauchy problem is 
   well-posed.
\end{itemize}

\textcolor{black}{
Returning to the specific case of the RCCZ4 formulation in vacuum, we 
linearize~(\ref{rccz4_chi_rccz4})--(\ref{rccz4_Zh_rccz4}) about some generic 
solution and consider perturbations in the high frequency regime. In such a 
regime, the length scale associated with the unperturbed solution will be large 
relative to the perturbations and we may safely freeze the coefficients in the 
perturbed equations. Upon decomposing the resulting linear constant coefficient 
into Fourier modes, we obtain:}
\begin{align}
\label{rccz4_hyperstart_rccz4_fo}
   \p_0 \chih &= -\frac{1}{6}\alpha \Kh + \frac{1}{6}\lp(i\omega_k\rp)
   \betah^k,
\\
   \p_0 \Kh &= \alpha \Rh +\omega_l\omega_m\g^{lm}\alphah +2\alpha\g^{lm} 
   \lp(i\omega_l
   \Zh_m\rp),
\\
   \p_0\hat{\Thetat} &= \frac{1}{2}\alpha^2 \lp( \Rh +2\lp(i\omega_i\rp)
   \Zh_j\g^{ij} \rp),
\\
   \p_0 \gth_{ij} &= -2 \alpha \Ath_{ij} -\frac{2}{3}\gt_{ij}
   \lp(i\omega_m\rp)
   \betah^m
   \\ \nonumber
   & \hphantom{=}
   + \gt_{im}\lp(i\omega_j\rp)\betah^{m} + \gt_{mj}\lp(i\omega_i\rp)
   \betah^{m},
\\
   \p_0 \Ath_{ij} &= e^{-4\chi}\lp[\omega_i\omega_j\g^{ij}\alphah +\alpha 
   \Rh_{ij}
   \rp.
   \\ \nonumber
   &\mathopen{}\lp.\vphantom{\frac{}{}}\hphantom{=}
   +2\alpha \lp( i\omega_{(i}\Zh_{j)} \rp)\rp]^{\mathrm{TF}},
\end{align}
\begin{align}
   \p_0\Lth^{i} &= \gt^{mn}\lp(-\omega_m\omega_n\rp)\betah^i 
   + \frac{1}{3}\gt^{ik}\lp(-\omega_k\omega_n\rp)\betah^n
   \\ \nonumber
   & \hphantom{=}
   -\frac{4}{3}\alpha\gt^{ij}\lp(i\omega_j\rp)\Kh 
   + 2\gt^{ik}\lp(i\omega_k\rp)\hat{\Thetat},
\\
   \p_0\Zh_i &= \alpha\lp[ \lp(i\omega_j\rp) \Ath_{ki}\gt^{jk} 
   -\frac{2}{3}\lp(i\omega_i\rp)\Kh \rp] + \lp(i\omega_j\rp)\hat{\Thetat},
\\
   \p_0 \alphah &= -\alpha^2 h \Kh + \alpha {h} {m} \hat{\Thetat},
\\
   \p_0 \betah^i &= \alpha^2 G \Bh^i,
\\
\label{rccz4_hyperend_rccz4_fo}
   \p_0 \Bh^i &= 2H\g^{im}\p_0\Zh_m + H\lp(i\omega^{n}\rp)\gt^{mi}
   \p_0\gth_{mn}.
\end{align}
\textcolor{black}{
Here, since we are interested in the high frequency regime, we have kept 
only the leading order derivative terms. In these equations, $\hat{R}_{ij}$ 
may either be considered as a function of $\Scale[1.0]{\Lamt}\vphantom{}^i$ 
(as would be the case for GBSSN):}
\begin{align}
   \Rh_{ij} &= \frac{1}{2}\gt^{lm}\lp(\omega_l\omega_m\rp)\gth_{ij} 
   + \frac{1}{2}\gt_{mi}\lp(i\omega_j\rp)\Lth^m 
   \\ \nonumber
   & \hphantom{=}
   +\frac{1}{2}\gt_{mj}\lp(i\omega_i\rp)\Lth^{m} 
   + 2\lp(\omega_i\omega_j\rp)\chih 
   \\ \nonumber
   & \hphantom{=}
   + 2\g_{ij}\g^{lm}\lp(\omega_l\omega_m\rp)\chih,
\end{align}
or as a function of $\Scale[1.0]{\Dt}\vphantom{}^i$ (as derived in 
Sec.~\ref{rccz4_sec_derivation}):
\begin{align}
   \Rh_{ij} &= \frac{1}{2}\gt^{lm}\lp(\omega_l\omega_m\rp)\gth_{ij} 
   + \frac{1}{2}\gt_{mi}\lp(i\omega_j\rp)\Dth^m 
   \\ \nonumber
   & \hphantom{=}
   +\frac{1}{2}\gt_{mj}\lp(i\omega_i\rp)\Dth^{m} 
   + 2\lp(\omega_i\omega_j\rp)\chih 
   \\ \nonumber
   & \hphantom{=}
   + 2\g_{ij}\g^{lm}\lp(\omega_l\omega_m\rp)\chih.
\end{align}
In what follows,  $\epsilon=1$ corresponds to the use of 
$\Scale[1.0]{\Dt}\vphantom{}^i$ while $\epsilon=2$ corresponds to the 
definition in terms of $\Scale[1.0]{\Lamt}\vphantom{}^i$. Roughly 
following~\cite{cao2022note}, we introduce the variables:
\begin{align}
   \omega_i &= \wabs\omegat_i,
\\
   \wabs^2 &= \g^{ij} \omega_i\omega_j,
\\
\label{rccz4_hypervar_alphah}
   \alphah &= \frac{-i \alpha}{\wabs}\ah,
\\
\label{rccz4_hypervar_chih}
   \chih &= \frac{-i}{\wabs}\Xh,
\\
\label{rccz4_hypervar_omegah}
\hat{\Thetat} &= \alpha\Omegah,
\\
\label{rccz4_hypervar_lambdah}
   \Lth^i &= \gt^{ij}\Lth_j,
\\
\label{rccz4_hypervar_betahh}
   \betah^i &= \frac{-i \alpha}{\wabs}\g^{ij}\bh_j,
\\
\label{rccz4_hypervar_Bh}
   \Bh^i &=\g^{ij}\Bh_{j},
\\
\label{rccz4_hypervar_gammath}
   \gth_{ij} &= \frac{-i e^{-4\chi}}{\wabs} \lh_{ij},
\\
\label{rccz4_hypervar_Ah}
   \Ath_{ij} &= e^{-4\chi}\LLh_{ij},
\end{align}
which permits us to 
write~(\ref{rccz4_hyperstart_rccz4_fo})--(\ref{rccz4_hyperend_rccz4_fo}) 
 as a first order pseudodifferential system of the form 
\begin{align}
   \label{rccz4_hyper_evo_uh}
   \p_0 \uh &= i\wabs\alpha \mathbf{P}\uh,
\end{align}
where
\begin{align}
   \uh &= 
   \begin{bmatrix}
      \ah & \chih & \Omegah & \Kh & \bh_i & \Bh_i & \Lth_i & \lh_{ij} 
      & \LLh_{ij}
   \end{bmatrix}^{\mathrm{T}}.
\end{align}
Provided that $\mathbf{P}$ is  diagonalizable with purely real eigenvalues,
the system will be strongly hyperbolic~\cite{nagy2004strongly, cao2022note, 
mongwane2016hyperbolicity}. Then, following the methodology of 
Nagy et al.~\cite{nagy2004strongly, cao2022note}, we decompose the eigenvalue 
equation
\begin{align}
\label{rccz4_hyperPuh_eqn}
   \mathbf{P}\uh = \lambda \uh,
\end{align}
by projecting $\uh$ into longitudinal and transverse components with respect
to $\omegat_i$ via application of the projection operator
\begin{align}
   q_{ij} &= \g_{ij} -\omegat_{i} \omegat_{j}.
\end{align}
Explicitly, we split all rank-1 and 2 covariant tensors into their components
in and orthogonal to $q_{ij}$. In such a decomposition, symmetric rank-2 
tensors on the 3D hypersurface with metric $\g_{ij}$ may be represented as:
\begin{align}
\label{rccz4_Xh_decomposition}
   \Xh_{ij} &= \omegat_i\omegat_j\Xh +\frac{1}{2}q_{ij}\Xhb 
   + 2\omegat_{(i}\Xhb_{j)} + \Xhb_{\langle ij \rangle},
\end{align}
with
\begin{align}
   \Xh &= \omegat^{i} \omegat^{j}\Xh_{ij},
\\
  \Xhb &= q^{ij} \Xh_{ij},
\\
   \Xhb_i &= {q_i}^{j}\omegat^{k}\Xh_{jk},
\\
   \Xhb_{\langle ij \rangle} &= {q_i}^{l}{q_j}^{m}\lp(\Xh_{lm} 
   -\frac{1}{2} \Xhb q_{lm}\rp),
\end{align}
and where angle brackets denote a tensorial quantity which is trace free with 
respect to $q_{ij}$. Similarly, covectors may be split according to
\begin{align}
\label{rccz4_Yh_decomposition}
   \Yh_{i} &= \omega_i\Yh + \Yhb_i,
\end{align}
with
\begin{align}
   \Yh &= \omegat^i\Yh_i, \\
   \Yhb_i &= {q_{i}}^{j}\Yh_j.
\end{align}

Upon application of these tensor and vector decompositions 
to~(\ref{rccz4_hyper_evo_uh}), we find that $\mathbf{P}$ can be written in 
block diagonal form:
\begin{align}
\label{rccz4_hyperP_matrix_decomposition}
   \mathbf{P} &= 
   \begin{bmatrix}
      \mathbf{P}^{\mathrm{S}} & 0 & 0 \\
      0 & \mathbf{P}^{\mathrm{V}} & 0 \\
      0 & 0 & \mathbf{P}^{\mathrm{T}}
   \end{bmatrix},
\end{align}
with $\Scale[1.0]{\mathbf{P}^{\mathrm{S}}}$, $\Scale[1.0]{\mathbf{P}
^{\mathrm{V}}}$ and $\Scale[1.0]{\mathbf{P}^{\mathrm{T}}}$ denoting scalar, 
vector and tensor components. Following a lengthy calculation, we find the 
following results for (1) the scalar components:
\begin{align}
\label{rccz4_pdr_rccz4_start}
   \p_0 \ah &= i \wabs\alpha \lp[ -h \Kh +{h} {m} \Omegah \rp],
\\
   \p_0 \bh &= i\wabs\alpha \lp[ G \Bh \rp],
\\
   \p_0 \Bh &= i\wabs\alpha \lp[ \frac{4H}{3}\bh -\frac{4H}{3}\Kh +2H\Omegah 
   \rp],
\\
   \p_0 \Xh &= i\wabs\alpha \lp[ \frac{1}{6}\bh -\frac{1}{6}\Kh  \rp],
\\
   \p_0 \lh &= i\wabs\alpha \lp[ \frac{4}{3}\bh -2\LLh \rp],
\\
   \p_0 \lhb &= i\wabs\alpha \lp[ -\frac{4}{3}\bh +2\LLh \rp],
\\
   \p_0 \Kh &= i\wabs\alpha \lp[-\ah -8\Xh + \frac{1}{2}\lh 
   -\frac{1}{2}\lhb +2\epsilon \Zh \rp],
\\
   \p_0 \Omegah &= i\wabs\alpha \lp[ -4\Xh + \frac{1}{4}\lh 
   -\frac{1}{4}\lhb +\epsilon\Zh \rp],
\\
   \p_0 \LLh &= i\wabs\alpha \lp[ -\frac{2}{3}\ah -\frac{4}{3}\Xh +
   \frac{1}{3} \lh +\frac{1}{6}\lhb +\frac{4\epsilon}{3}\Zh  \rp],
\\
   \p_0\Zh &= i\wabs\alpha\lp[ \LLh -\frac{2}{3}\Kh + \Omegah \rp],
\end{align}
(2) the vector components:
\begin{align}
   \p_0 \bhb_i &= i\wabs\alpha \lp[ G \Bhb_i \rp],
\\
   \p_0 \Bhb_i &= i\wabs\alpha \lp[ H \bhb_i \rp],
\\
   \p_0 \lhb_i&= i\wabs\alpha \lp[ \bhb_i -2\LLhb_i \rp],
\\
   \p_0 \LLhb_i &= i\wabs\alpha \lp[ \epsilon\Zhb_i  \rp],
\\
   \p_0 \Zhb_i &= i\wabs\alpha \lp[ \LLhb_i \rp],
\end{align}
and (3) tensor components:
\begin{align}
   \p_0 \lhb_{\langle ij \rangle} &= i\wabs\alpha 
   \lp[ -2 \LLhb_{\langle ij  \rangle} \rp],
\\
\label{rccz4_pdr_rccz4_end}
   \p_0 \LLhb_{\langle ij \rangle} &= i\wabs\alpha
   \lp[ -\frac{1}{2}\lhb_{\langle ij  \rangle}  \rp].
\end{align}
Note that since $\Ah_{ij}$ is trace-free we have $\LLhb = -\LLh$,
which is why no evolution equation for $\LLhb$ appears. Expressing these 
systems of equations as matrix equations of the 
form~(\ref{rccz4_hyperPuh_eqn}) and 
(\ref{rccz4_hyperP_matrix_decomposition}), the eigenvalues of 
$\Scale[1.0]{\mathbf{P}^{\mathrm{S}}}$ are:
\begin{align}
   \lambda &=0,\,0,\,\pm 1,\, \pm\sqrt{\epsilon},\, \pm\sqrt{h},\, \pm 
   \sqrt{\frac{4}{3}GH}.
\end{align}
Comparing with the results of \cite{bernuzzi2010constraint, cao2022note} 
(which consider various BSSN-type systems), we observe that RCCZ4 has one 
fewer zero velocity eigenvalue than GBSSN. It is this eigenvalue which 
corresponds to the Hamiltonian constraint advection and it is largely 
responsible for the superior performance of FCCZ4 relative to 
GBSSN~\cite{daverio2018apples, bernuzzi2010constraint, hilditch2013compact}. 
Treating $\Rt_{ij}$ as a function of $\Lamt^i$ versus $\Dt^i$ ($\epsilon=2$ 
versus $\epsilon=1$) has the effect of increasing the speed of propagation 
of several modes, but otherwise has no effect on hyperbolicity. In fact, 
we see that RCCZ4 appears to be well defined for a fairly wide range of 
$\epsilon$ which roughly corresponds to modified equations of motion in 
which the Ricci tensor is supplemented by additional terms of the form 
$\dt_{(i}\zh_{j)}$. 

In the case of the vector components, the eigenvalues of the matrix 
$\Scale[1.0]{\mathbf{P}^{\mathrm{V}}}$ each have multiplicity 2 (rather than 
3) due to the projection constraints of the form $\omegat^i\Xh_i = 0$. The 
eigenvalues are: 
\begin{align}
   \lambda &=0,\, \pm \sqrt{\epsilon},\, \pm \sqrt{GH}.
\end{align} 
Finally, for the tensor components, the eigenvalues of $\Scale[1.0]{\mathbf{P}
^{\mathrm{T}}}$ have multiplicity 2 (rather than 6) due to the three 
projection constraints of the form $\omegat^i\Xhb_{ij} = 0$ and the 
trace-free condition $\Xhb_{\langle ij \rangle} \g^{ij} = 0$. The eigenvalues 
are the same as we would find for BSSN and ADM~\cite{cao2022note, 
nagy2004strongly, bernuzzi2010constraint}:
\begin{align}
   \lambda &=\pm 1.
\end{align}
In order to guarantee weak hyperbolicity, all of these eigenvalues must be 
real, so we must have 
\begin{align}
   GH > 0, \; h > 0, \; \epsilon > 0. 
\end{align}
Strong hyperbolicity additionally requires that each of $\mathbf{P}^S$, 
$\mathbf{P}^V$ and $\mathbf{P}^T$ are diagonalizable. For this to be the
case, all of the following conditions must hold: 
\begin{align}
   h \neq \epsilon, \;
   HG \neq \frac{3}{4}, \;
   HG \neq \frac{3}{4}h, \;
   HG \neq \frac{3}{4}\epsilon,
\end{align}
so that $\mathbf{P}$ has a complete set of eigenvectors.
\textcolor{black}{
Here, $\epsilon\notin \{1,2\}$ would occur if we were to substitute some other 
combination of $\Dt^i$ and $\gt^{ij}\bar{Z}_j$ in the definition of $\Rt_{ij}$. 
Note that as $h$, $G$ and $H$ are generically functions of $\alpha$ and 
$\chi$, we cannot guarantee that our equations of motion will be everywhere 
strongly hyperbolic. However, if we perform the same sort of 
pseudodifferential decomposition for FCCZ4 (using a slightly modified gauge), 
we find that RCCZ4 and FCCZ4 share the same principal part and we thus 
conclude that the two methods have identical stability characteristics in 
the high frequency limit.}

\section{\label{rccz4_sec_conclusions}Summary and Conclusions}
In this paper,
we have introduced our novel RCCZ4 formulation of numerical relativity. We 
have demonstrated that it is possible to achieve roughly equivalent 
performance to GBSSN and FCCZ4 through a modification of Z4 wherein 
constraint violations are coupled to a reference metric completely 
independent of the physical metric. We have shown that this approach works 
in the presence of black holes and holds up robustly in a variety of 1D 
simulations including the critical collapse of a scalar field. In addition 
to stably evolving spherically symmetric simulations in the strong field, we have 
demonstrated that our formulation is strongly hyperbolic through the use of 
a pseudodifferential first order reduction.

Our formulation of RCCZ4 chose the simplest possible reference metric, but
we can easily imagine formulations in which the components of 
$\accentset{\circ}{g}_{\mu \nu}$ are chosen or evolved  in such a way so 
as to provide additional beneficial properties aside from the vanishing of 
the Ricci tensor. We suspect that it will be in modifications to the choice 
of $\accentset{\circ}{g}_{\mu \nu}$ in which the full utility of RCCZ4-like 
formulations is realized.

The core idea behind RCCZ4---coupling the constraint equations to a metric
different from the physical metric---could potentially be used to derive
methods with greater stability and superior error characteristics than either
GBSSN or FCCZ4. In our opinion, the main takeaway should not be that RCCZ4, 
as it stands, is a complete formulation with performance approaching or 
exceeding FCCZ4 and GBSSN. 
Rather, the main lesson should be that the Z4 formulation of general 
relativity can be modified such that the constraints are coupled to a metric 
other then the physical one, and that such a modification may be useful in 
tailoring the properties of the system as they pertain to constraint 
advection and damping.

\begin{acknowledgments}
This research was supported by the Natural Sciences and Engineering Research 
Council of Canada (NSERC). 
\end{acknowledgments}

\appendix
% last edit 2023/09/05 at 5:06 PM

% GDR: sorry for not having gone through this between the MWC and MWC2 edits. 
% Sorry for the frustration that caused.
% GDR: pass complete
% GDR: replaced Z_i with \zh_i

% MWC3: There may be too much detail in some of the derivations here for 
% a publication, but let's wait until we hear what the referee has to say.
% GDR3: ok, it should be easy to trim them down anyway.

\section{3+1 Form of RZ4\label{app_rccz4_rz4_derivation}}

%In this appendix we provide a brief overview of the derivation of 
%(\ref{rccz4_gamma_ij})--(\ref{rccz4_z_i}), the ADM equivalent of RZ4. 
The RZ4 equations in canonical and trace-reversed form with damping are 
given by (\ref{rccz4_RZ4_v2}) and (\ref{rccz4_RZ4_v1}). As we have been 
predominantly interested in investigating scale invariant problems, we set 
the damping parameters $\kappa_1$ and $\kappa_2$ to zero, yielding the 
simpler set of equations:
\begin{align}
   \label{app_rccz4_RZ4_v2_simp}
   % verified (black notes, accz4 (eqn. 3))
   &R_{\mu\nu} -\frac{1}{2}g_{\mu\nu} R + 2\no_{(\mu}Z_{\nu)} 
   -g_{\mu\nu}\no_{(\alpha} Z_{\beta)} g^{\alpha\beta} 
   \\ \nonumber
   & \hphantom{=}
   -8\pi T_{\mu\nu}= 0,
\\
   \label{app_rccz4_RZ4_v1_simp}
   % verified (black notes, accz4 (eqn. 1))
   &R_{\mu\nu} + 2\no_{(\mu}Z_{\nu)}
   - 8\pi \lp( T_{\mu\nu}  -\frac{1}{2}g_{\mu\nu}T \rp) =0,
\\
   \label{app_rccz4_RZ4_trace_simp}
   % verified (black notes, accz4 (eqn. 2))
   &R + 2\no_{(\mu} Z_{\nu)}g^{\mu\nu} +8\pi T =0.
\end{align}

Here, (\ref{app_rccz4_RZ4_v2_simp}) is RZ4 written in canonical form, 
(\ref{app_rccz4_RZ4_v1_simp}) is written in trace-reversed form and  
(\ref{app_rccz4_RZ4_trace_simp}) is the trace of 
(\ref{app_rccz4_RZ4_v1_simp}) taken with respect to the physical metric 
$g^{\mu\nu}$.

To derive the ADM equivalent of the RZ4 equations we roughly follow the 
ADM derivations of~\cite{alcubierre2008introduction, gourgoulhon2012} and 
take projections of 
(\ref{app_rccz4_RZ4_v2_simp})--(\ref{app_rccz4_RZ4_trace_simp}) onto and 
orthogonal to the spatial hypersurfaces which foliate four dimensional 
spacetime in a standard {3+1} decomposition. In what follows, we consider 
% MWC7: Again, check this
% only the simplest case where $\accentset{\circ}{g}_{ij}$ is a 
% time-invariant, curvature-free Lorentzian metric with 
% $\accentset{\circ}{g}_{tt} = 1, \accentset{\circ}{g}_{tj} = 0$.
only the simplest case where $\accentset{\circ}{g}_{\mu\nu}$ is a 
time-invariant, curvature-free, Lorentzian metric with 
$\accentset{\circ}{g}_{tt} = -1, \accentset{\circ}{g}_{tj} = 0$.

\subsection{Spatial projection}

We begin by finding the evolution equation for the extrinsic curvature by 
projecting both indices of~(\ref{app_rccz4_RZ4_v1_simp}) onto $\Sigma$. The 
terms present in the Einstein equations follow the ordinary ADM derivation
so we concentrate on the terms containing $Z^\mu$:
% explicit check (4D): full_cart_check_2023_09_05: (174)
% explicit check: full_cart_check_2023_09_05: (399)-(404)
\begin{align}
% verified (black notes, accz4 (1))
        \g^{\mu}\vp_\lambda \g^{\nu}\vp_\sigma\no_\mu Z_\nu
        &= \g^{\mu}\vp_\lambda \g^{\nu}\vp_\sigma \lp(\vphantom{\frac{}{}} 
        \p_\mu {\zh}_\nu +\Theta \p_\mu n_\nu 
        \rp.
        \\ \nonumber
        &\mathopen{}\lp. \hphantom{=}
        - \Go^\rho \vp_{\mu \nu} \lp( {\zh}_\rho + n_\rho \Theta \rp) \rp).
\end{align} 
We now note that, since $n_{i} = 0$, when restricting to spatial indices we 
have:
% check explicit:  full_cart_check_2023_09_05 (174, commented out)
\begin{align}
% verified (black notes, accz4 (1))
        \g^{\mu}\vp_l \g^{\nu}\vp_m \p_\mu n_\nu &= 
        \lp({\delta^{\mu}}_l + n^{\mu}n_{l}\rp)
        \lp({\delta^{\nu}}_m + n^{\nu}n_{m}\rp)\p_{\mu}n_{\nu},
        \\ \nonumber
        &=\lp({\delta^{\mu}}_l{\delta^{\nu}}_m 
        + {\delta^{\mu}}_l n^{\nu}n_m
        + {\delta^{\nu}}_m n^{\mu}n_{l} 
        \rp.
        \\ \nonumber
        &\mathopen{}\lp. \hphantom{=}
        + n^{\mu}n_{l}n^{\nu}n_{m} \rp) \p_{\mu}n_{\nu},
        \\ \nonumber
        &=\lp( \p_{l} n_{m} + n^{\nu}n_m\p_{l}n_{\nu}
        + n^{\mu}n_{l}\p_{\mu}n_{m} 
        \rp.
        \\ \nonumber
        &\mathopen{}\lp.\hphantom{=}
        + n^{\mu}n_{l}n^{\nu}n_{m} \p_{\mu}n_{\nu} \rp),
        \\ \nonumber
        &=0,
\end{align}
and therefore
\begin{align}
% verified (black notes, accz4 (1))
        2\g^{\mu}\vp_i \g^{\nu}\vp_j\no_{(\mu} Z_{\nu)}
        &= 2\g^{\mu}\vp_i \g^{\nu}\vp_j \lp(\p_{(\mu} {\zh}_{\nu)} 
        +\Theta \p_{(\mu} n_{\nu)} 
        \rp.
        \\ \nonumber
        &\mathopen{}\lp.\hphantom{=}
        - \Go^\rho \vp_{\mu \nu} \lp({\zh}_\rho 
        + n_\rho \Theta \rp) \rp),
        \\  \nonumber
        &= 2\g^{\mu}\vp_i \g^{\nu}\vp_j \lp( \p_{(\mu} {\zh}_{\nu)} 
        - \Go^\rho \vp_{\mu \nu} \lp({\zh}_\rho 
        + n_\rho \Theta \rp) \vphantom{\frac{}{}}\rp).
\end{align} 
Assuming $\accentset{\circ}{g}_{\mu\nu} = \delta_{tt} +\go_{ij}$ with 
$\Ro_{ij} = {\Go^t}_{ij} = 0$ (e.g. we take the simplest possible flat 
background 3-metric), this simplifies further to,
\begin{align}
% verified (black notes, accz4 (5))
\label{app_rccz4_nabla_v_z_v_sigma_project}
        2\g^{\mu}\vp_i \g^{\nu}\vp_j\no_{(\mu} Z_{\nu)} 
        =& 2\g^{\mu}\vp_i \g^{\nu}\vp_j \lp( \p_{(\mu} {\zh}_{\nu)} 
        - \Go^k \vp_{\mu \nu} {\zh}_k \rp),
        \\  \nonumber
        =& 2\lp(\delta^\mu\vp_i \delta^\nu\vp_j + 
        \delta^\mu\vp_i n^\nu n_j +\delta^\nu\vp_j n^\nu n_i 
        \rp.
        \\ \nonumber
        &\mathopen{}\lp.
        + n^\mu n_i n^\nu n_j\rp)
        \lp(\p_{(\mu} {\zh}_{\nu)} 
        - \Go^k \vp_{\mu \nu} {\zh}_k \rp),
        \\  \nonumber
        =& 2\p_{(i} {\zh}_{j)} 
        - 2\Go^k \vp_{i j} {\zh}_k,
        \\  \nonumber
        =& 2\Do_{(i} {\zh}_{j)} \, .
\end{align}
Here, we have made use of the fact that, with the connection given above, the 
Christoffel symbols for the spatial component of the background metric are 
identical to those of its four dimensional counterpart. Adding 
(\ref{app_rccz4_nabla_v_z_v_sigma_project}) to the evolution equation 
for the extrinsic curvature,
\begin{align}
   \label{app_rccz4_evo_K_adm}
   \lm K_{ij}&= -D_i D_j \alpha + \alpha
   \lp(R_{ij} + K K_{ij} - 2 K_{ik}K^{k}\vp_j\rp) 
   \\ \nonumber
   & \hphantom{=}
   +4\pi\alpha\lp(\g_{ij}\lp(S-\rho\rp)-2 S_{ij}\rp),
\end{align}
we recover~(\ref{rccz4_k_ij}).

\subsection{Temporal Projection}

Next, we modify the Hamiltonian constraint by considering the full 
projection of~(\ref{app_rccz4_RZ4_v2_simp}) onto $n^\mu n^\nu$.  Focusing on 
the terms that have been added to the original Einstein equations we have:
\begin{align}
% verified (black notes, accz4 (2))
        &
        n^\mu n^\nu \no_\mu Z_\nu 
        %&
        \\ \nonumber
        & \hphantom{=}
        = n^\mu \no_\mu \lp(n^\nu Z_\nu\rp) 
        - n^\mu Z_\nu \no_\mu n^\nu ,
        \\  \nonumber
        & \hphantom{=}
        = -n^\mu \no_\mu \Theta - n^\mu Z_\nu \no_\mu n^\nu,
        \\  \nonumber
        & \hphantom{=}
        = -\frac{1}{\alpha}\lm \Theta - n^\mu Z_\nu \no_\mu n^\nu,
\\
% verified (black notes, accz4 (2))
        &
        n^\mu n^\nu g_{\mu\nu} \lp(\no_\lambda Z_\sigma\rp) g^{\lambda\sigma}
        %&
        \\ \nonumber
        & \hphantom{=}
        = - \lp(\no_\lambda Z_\sigma\rp) g^{\lambda \sigma},
        \\  \nonumber
        & \hphantom{=}
        =-g^{\lambda\sigma}\no_\lambda\lp({\zh}_\sigma +n_\sigma\Theta\rp),
        \\  \nonumber
        & \hphantom{=}
        =-\frac{1}{\alpha}\lm\Theta -g^{\lambda\sigma} 
        \no_\lambda {\zh}_\sigma 
        %\\ \nonumber
        %& \hphantom{=}
        % \hphantom{=}
        -g^{\lambda \sigma}\Theta\no_\lambda n_\sigma.
\end{align}
Thus, we find
\begin{align}
% verified (black notes, accz4 (2))
\label{app_rccz4_temporal_project_1}
        &
        n^\mu n^\nu \lp(2\no_{(\mu} Z_{\nu)} -g_{\mu\nu} \no_\lambda Z_\sigma 
        g^{\lambda \sigma} \rp)
        %&
        \\ \nonumber
        &\hphantom{=}
        = -\frac{1}{\alpha} 
        \lm\Theta -2 n^\mu\lp( {\zh}_\nu +n_\nu \Theta \rp) 
        \no_\mu n^\nu 
        \\ \nonumber
        & \hphantom{==}
        +g^{\lambda \sigma}\no_\lambda {\zh}_\sigma 
        + g^{\lambda \sigma} \Theta \no_\lambda n_\sigma.
\end{align}
% MWC:  Be explicit about the substitution and note that two equations
% are of comparable complexity
% GDR: changed
Now, expressing $n^{\mu}$ and $g_{\mu\nu}$ in terms of $\alpha$, $\beta^i$ 
and $\g_{ij}$ and simplifying, (\ref{app_rccz4_temporal_project_1}) 
becomes:
% explicit check: full_cart_check_2023_09_05 (406)
\begin{align}
% verified (black notes, accz4 (6.3))
        &
        n^\mu n^\nu \lp(2\no_{(\mu} Z_{\nu)} -g_{\mu\nu} \no_\lambda Z_\sigma 
        g^{\lambda \sigma} \rp)
        %&
        \\ \nonumber
        &\hphantom{=}
        =
        -\frac{1}{\alpha} \lm\Theta 
        - \frac{\Theta}{\alpha^2} \lm \alpha
        + \frac{{\zh}_i}{\alpha^2} \lp(\lm\beta^i - \beta^j\Do_j \beta^i\rp) 
        \\ \nonumber
        & \hphantom{==}
        %\\ \nonumber
        %& \hphantom{=}
        + \g^{ij}\Do_i {\zh}_j.
\end{align}
Adding these to the ADM Hamiltonian constraint,
\begin{align}
   H &= \frac{1}{2}\lp(R + K^2 - K_{ij}K^{ij}\rp) - 8\pi\rho = 0, 
\end{align}
and solving for $\lm \Theta$, we 
recover~(\ref{rccz4_theta}). 

\subsection{Mixed Projection}

We find the evolution equation for the momentum constraint propagator by 
taking the mixed projection onto $\g^\mu\vp_\lambda n^\nu$ of the terms 
that have been added to the Einstein equations in 
(\ref{app_rccz4_RZ4_v2_simp}). Upon restricting to spatial indices we find:
\begin{align}
% verified (black notes, accz4 (3))
        &
        \g^\mu\vp_i n^\nu \no_\mu Z_\nu 
        &
        \\ \nonumber
        & \hphantom{=}
        = -\g^\mu\vp_i \no_\mu \Theta
        -\g^\mu\vp_i Z_\nu \no_\mu n^\nu,
        \\ \nonumber 
        &\hphantom{=}
        =-\p_i \Theta -n_i n^\mu \no_\mu \Theta - \g^{\mu}\vp_i \lp(
        {\zh}_\nu 
        + \Theta n_\nu\rp) \no_\mu n^\nu,
        \\ \nonumber
        &\hphantom{=}= -D_i \Theta - \g^{\mu}\vp_i \lp(
        {\zh}_\nu + \Theta n_\nu\rp) \no_\mu n^\nu,
\\
% verified (black notes, accz4 (3))
        &
        \g^\mu\vp_i n^\nu \no_\nu Z_\mu 
        %&
        \\ \nonumber
        &\hphantom{=}
        = n^\nu \no_\nu \lp(\g^{\mu}\vp_i Z_\mu\rp)
        -Z_\mu n^\nu \no_\nu \g^{\mu}\vp_i,
        \\ \nonumber
        &\hphantom{=}
        = n^\nu \no_\nu {\zh}_i -Z_\mu n^\nu \no_\nu \lp( \delta^\mu\vp_i
        +n^\mu n_i \rp) ,
        \\ \nonumber
        &\hphantom{=}
        = n^\nu \no_\nu {\zh}_i -n_i Z_\mu n^\nu \no_\nu n^\mu
        + \Theta n^\nu \no_\nu n_i,
        \\ \nonumber
        &\hphantom{=}
        = n^\nu \no_\nu {\zh}_i + \Theta n^\nu \no_\nu n_i,
        \\ \nonumber
        % GDR: check especially: (double check that the Theta term 
        % goes away)
        &\hphantom{=}
        = \frac{1}{\alpha} \lm {\zh}_i - {\zh}_\mu \no_i n^\mu,
\\
% verified (black notes, accz4 (4))
        &
        \g^\mu\vp_i n^\nu \lp( g_{\mu\nu} g^{\lambda \sigma} 
        \no_\lambda Z_\sigma \rp)
        %& 
        \\ \nonumber 
        & \hphantom{=}
        =\g^\mu\vp_i n^\nu \lp( \g_{\mu\nu} -n_\mu n_\nu\rp) 
        \lp(g^{\lambda \sigma} \no_\lambda Z_\sigma \rp),
        \\ \nonumber
        & \hphantom{=}
        = 0.
\end{align}
% MWC: Again, references to specific equations are called for
% GDR: made it explicit what is being expressed in terms of what.
Now, expressing $n^{\mu}$ and $g_{\mu\nu}$ in terms of the 3+1 variables 
($\alpha$, $\beta^i$ and $\g_{ij}$) and simplifying the resulting 
expression, we find:
\begin{align}
% checked explicit: full_cart_check_2023_09_05: (215)-(217)
% verified (black notes, accz4 (8))
   &
   \g^\mu\vp_a n^\nu \lp( 2 \no_{(\mu}Z_{\nu)} -g_{\mu\nu} 
   g^{\lambda\sigma} \no_l Z_m \rp) 
   %&
   \\ \nonumber
   &\hphantom{=}
   =\frac{1}{\alpha} \lm {\zh}_i - {\zh}_\mu \no_i n^\mu -\Do_i \Theta
   \\ \nonumber
   & \hphantom{==}
   - \g^{\mu}\vp_i \lp({\zh}_\nu + \Theta n_\nu\rp) \no_\mu n^\nu,
   \\ \nonumber
   & \hphantom{=}
   = \frac{1}{\alpha} \lm {\zh}_i + \frac{2}{\alpha} {\zh}_j \Do_i 
   \beta^j 
   -\Do_i\Theta - \Theta \Do_i\lnalpha.
\end{align}
Upon substitution of this expression into
the ADM momentum constraint,
\begin{align}
   \label{app_rccz4_mom_adm}
   M^{i} &= D_{j}K^{ij} - \gamma^{ij}D_{j} K -8\pi j^{i} = 0,
\end{align} 
and solving for $\lm {\zh}_i$, we recover~(\ref{rccz4_z_i}).

\section{Derivation of RCCZ4\label{app_rccz4_rccz4_derivation}}

Now that we have the ADM equivalent of the RZ4 equations, the derivation of 
the RCCZ4 equations proceeds in a fairly straightforward manner. To recap, 
the ADM equivalents of the RZ4 equations so far derived are:
% verified all in maple full_cat_check_2023_09_05 (around 800)
\begin{align}
% verified (notes)
\label{app_rccz4_gamma_ij}
   \lm\g_{ij} &= -2\alpha K_{ij},    
\\
% verified (notes)
\label{app_rccz4_k_ij}
   \lm K_{ij} &= -D_i D_j \alpha +\alpha \lp( R_{ij} + K K_{ij} 
   -2 K_{ik} K^{k}\vp_{j}\rp)
   \\ \nonumber
   & \hphantom{=}
   +4\pi\alpha \lp( \lp[ S-\rho\rp]\g_{ij} -2S_{ij} \rp) + 2 \alpha
   \Do_{(i} \zh_{j)},
%\\
\end{align}
\begin{align}
% verified (notes)
\label{app_rccz4_theta}
   \lm \Theta &= \frac{\alpha}{2}\lp(R + K^2 - K_{ij}K^{ij}
   - 16\pi\rho \rp) 
   \\ \nonumber
   & \hphantom{=}
   + \alpha\g^{ij}\Do_i \zh_{j} -\frac{\Theta}{\alpha}\lm \alpha 
   \\ \nonumber
   & \hphantom{=}
   +\frac{\zh_i}{\alpha}\lp(\lm \beta^i - \beta^j\Do_j\beta^i\rp),
\\
% verified
\label{app_rccz4_z_i}        
   \lm \zh_i &= \alpha \lp( D_j K^{j}\vp_i -D_i K -8\pi j_i \rp) 
   -2 \zh_j\Do_i\beta^j 
   \\ \nonumber
   & \hphantom{=}
   +\Theta\Do_i\alpha +\alpha\Do_i\Theta,
\end{align} 
and the process of determining the RCCZ4 equations essentially boils down 
to substituting for the conformal variables in a manner exactly analogous to 
FCCZ4~\cite{sanchis2014fully}.

We observe that~(\ref{app_rccz4_gamma_ij}), the evolution equation for 
$\g_{ij}$, is unchanged from the ADM case and therefore the evolution 
equations for $\chi$ and $\gh_{ij}$ are the same as in FCCZ4 and 
GBSSN~\cite{sanchis2014fully, brown2009covariant}:
\begin{align}
% verified (notes)
\label{app_rccz4_chi_rccz4}
        \lm \chi &= -\frac{1}{6} \alpha K + \frac{1}{6}  \dt_m\beta^m,
\\
% verified (notes)
\label{app_rccz4_gh_rccz4}
        \lm \gt_{ij} &= -2 \alpha \At_{ij} - \frac{2}{3} \gt_{ij} \dt_{m}
        \beta^{m}.
\end{align}

%\subsection[Evolution of the Extrinsic Curvature Trace]
%{Evolution of $K$}
\subsection{Evolution of the Extrinsic Curvature Trace}

Beginning with the Lie derivative of $K$ along $m$:
\begin{align}
        \lm K &= \g^{ij}\lm K_{ij} + K_{ij}\lm \g^{ij},
\end{align}
and (\ref{app_rccz4_k_ij}), the RZ4 form of the evolution of the extrinsic curvature, we substitute (\ref{app_rccz4_gamma_ij}) for $\lm \g^{ij}$, to find~(\ref{rccz4_K_rccz4}):
\begin{align}
% verified black notes (Z4 redo)
\label{app_rccz4_lm_K_rccz4}
   \lm K &= \g^{ij}\lm K_{ij} + K_{ij}\lm \g^{ij}
   \\ \nonumber
   &
   = -D^i D_i \alpha +\alpha \lp( R + K^2 -2 K_{ij} K^{ij}\rp)
   \\ \nonumber
   & \hphantom{=}
   +4\pi\alpha \lp(3 \lp[ S-\rho\rp] -2S \rp) + 2 \alpha \g^{ij} 
   \Do_{(i} \zh_{j)} 
   \\ \nonumber 
   &\hphantom{=}
   +2\alpha K_{ij}K^{ij}
   \\ \nonumber
   &=-D^2\alpha +\alpha \lp( R +K^2 +2\g^{ij} \Do_{(i} {\zh}_{j)} 
   \rp.
   \\ \nonumber
   &\lp.\vphantom{\frac{}{}}\hphantom{=}
   +4 \pi \lp(S-3\rho\rp) \rp).
\end{align}

%\subsection[Evolution of the Trace-Free Extrinsic Curvature]
%{Evolution of $\A_{ij}$}
\subsection{Evolution of the Trace-Free Extrinsic Curvature}
% GDR: in the process of checking something with this derivation...
% did I make a substitution of the Hamiltonian constraint?
% No, I did not, everything is fine, the derivation just has an
% annoying number of parts and I neglected a factor of 2/3 in my
% check which messed everything up...
The evolution of $\lm \At_{ij}$ is given by
\begin{align}
   \lm \At_{ij} &= \lm\lp(\e4c\lp( K_{ij} - \frac{1}{3}\g_{ij} K \rp)\rp)
   \\ \nonumber
   &= -4 \At_{ij} \lm \chi +\e4c\lp( \lm K_{ij} -\frac{1}{3}K\lm \g_{ij}
   \rp.
   \\ \nonumber
   &\lp. \hphantom{=} 
   -\frac{1}{3} \g_{ij} \lm K \rp) .
\end{align}
If we express this equation in terms of the conformal decomposition and make 
use of (\ref{app_rccz4_k_ij}) and (\ref{app_rccz4_lm_K_rccz4}), the RZ4 
evolution equations for $K_{ij}$ and $K$ respectively, we 
find~(\ref{rccz4_ah_rccz4}):
% verified (Black notes (9 & 15))
\begin{align}
   \lm \At_{ij} &= \e4c\lp[ -D_i D_j \alpha +\alpha R_{ij} 
   - 8\pi\alpha S_{ij} 
   \vphantom{\frac{}{}}
   \rp.
   \\ \nonumber
   &\lp. \hphantom{=}
   +2\alpha\Do_{(i} \zh_{j)} \rp]^\tf +\alpha\lp(K \At_{ij} 
   -2 \At_{ik} {\At^{k}}_j\rp) 
   \\ \nonumber
   & \hphantom{=}
   -\frac{2}{3}
   \At_{ij}\dt_l \beta^{l}.
\end{align}

Equivalently, we could start from the GBSSN equation for 
$\At_{ij}$~\cite{brown2009covariant, alcubierre2011formulations}:
\begin{align}
   \label{app_rccz4_lm_A_ij_adm}
   \lm \At_{ij} &= \e4c\lp[ -D_i D_j \alpha +\alpha R_{ij} 
   - 8\pi\alpha S_{ij}  \rp]^\tf 
   \\ \nonumber
   & \hphantom{=}
   +\alpha\lp(K \At_{ij} -2 \At_{ik} {\At^{k}}_j\rp) 
   -\frac{2}{3} \At_{ij}\dt_l \beta^{l},
\end{align}
and note that~(\ref{app_rccz4_k_ij}) is, save for the term involving 
$\Do_{(i} \zh_{j)}$, identical to the ADM expression for the evolution for 
the extrinsic curvature. If we define 
% MWC3: Don't use mathbb
% GDR3: ok, replacing with bar
\begin{align}
        \bar{R}_{ij} = R_{ij} +2 \Do_{(i}{\zh}_{j)},
\end{align}
and note that this new pseudo-curvature has the same symmetries as a 
true curvature, we may follow the \mbox{GBSSN} derivation of $\lm \Ah_{ij}$ 
exactly  and substitute the definition of this new quantity as a final step. 
Doing so recovers~(\ref{rccz4_ah_rccz4}) in a much simpler manner.

%\subsection[Evolution of Theta]
%{Evolution of $\Theta$}
\subsection{Evolution of Theta}

Essentially trivial substitution of the conformal variables into 
(\ref{app_rccz4_theta}), the augmented Hamiltonian constraint, 
gives:
\begin{align}
\label{app_rccz4_lm_theta_simp}
   \lm \Theta &= \frac{\alpha}{2}\lp(R - \Ah_{ij}\Ah^{ij} +\frac{2}{3}K^2
   - 16\pi\rho \rp) 
   \\ \nonumber
   & \hphantom{=}
   + \alpha\g^{ij}\Do_i \zh_{j} -\frac{\Theta}{\alpha}\lm \alpha 
   %\\ \nonumber
   %& \hphantom{=}
   +\frac{\zh_i}{\alpha}\lp(\lm \beta^i - \beta^j\Do_j\beta^i\rp).
\end{align}

%\subsection[Evolution of Lambda]
%{Evolution of $\Lh^i$}
\subsection{Evolution of Lambda}
From~(\ref{rccz4_Lamt_def}), the definition of $\Lamt^i$ we find the 
following expression for the evolution of $\lm \Lamt^i$
\begin{align}
% verified (black notes accz4 11)
% GDR: lets check this in maple since it is an intermediary equation: error
\label{app_rccz4_lm_lamt_simp}
   \lm{\Lamt^i} &= \lm{\Dt^i} + 2\lm{\lp(\gt^{ij} \zh_j\rp)}.
\end{align}
In this equation, an expression for $\lm \zh_i$ may be found through 
substitution of the conformal variables into (\ref{app_rccz4_z_i}):
\begin{align}
   \lm \zh_i &= \alpha \lp( D_l \Ah^{l}\vp_i -\frac{2}{3}D_i K -8\pi j_i \rp) 
   -2 \zh_j\Do_i\beta^j 
   \\ \nonumber
   & \hphantom{=}
   +\Theta\Do_i\alpha +\alpha\Do_i\Theta.
\end{align}

Now, the quantity $\Dt^i$ can be expressed in terms of the action of the 
flat space covariant derivative on the conformal metric:
% verified (black notes (18-22))
\begin{align}
   \Do_j \gt^{ij} = -\Dt^i - \frac{1}{2} D_k \ln{\lp(\frac{\gt}{\go}\rp)}\gt^{ik},
\end{align}
and, noting that since $\gt=\go$ (we have chosen our conformal and flat space 
metrics to have the same determinant), $\Dt^i$ may be expressed as:
\begin{align}
   \Dt^i &= -\Do_j \gt^{ij}.
\end{align}
We may then find an evolution equation for $\Dt^i$ entirely in terms 
of~(\ref{app_rccz4_gh_rccz4}), the equation of motion for $\gt_{ij}$, and 
the definition of ${\Dt^{i}}_{jk}$:
\begin{align}
% verified (black notes, GBSSN 18-22 (3))
   \lm \Dh^i &= \gh^{mn}\Do_m\Do_n\beta^i -2\Do_j\lp(\alpha\Ah^{ij}\rp)
   \\ \nonumber
   & \hphantom{=}
   +\frac{1}{3}\gh^{mi}\Do_m\Do_n\beta^n + \frac{2}{3}\Dh^i\Do_m\beta^m.
\end{align}

Finally, (\ref{app_rccz4_lm_lamt_simp}) may be expressed as:
\begin{align}
% verified (black notes accz4 11)
% GDR: lets check this in maple since it is an intermediary equation:
% checked in full_cart_check_2023_09_05
   \lm{\Lamt^i} &= \lm{\Dt^i} + 2\lm{\lp(\gt^{ij} \zh_j\rp)},
   \\ \nonumber 
   &= \gh^{mn}\Do_m\Do_n\beta^i -2\Ah^{ij}\Do_j\alpha 
   +\frac{1}{3}\gh^{mi}\Do_m\Do_n\beta^n
   \\ \nonumber
   & \hphantom{=}
   +\frac{2}{3} \Lamt^i\dt_n\beta^n + 4\alpha\zh_j\At^{ij} 
   +12\alpha\At^{li}\Dt_l\chi 
   \\ \nonumber
   & \hphantom{=}
   -\frac{4}{3}\alpha\dt^i K -16\pi\alpha\jt^i + 2\alpha\dt^i\Theta 
   +2\alpha\Theta\dt^i\ln{\alpha} 
   \\ \nonumber
   & \hphantom{=}
   - 4 \zh_l \gt^{ij}\Do_j \beta^l .
\end{align}

\subsection{Simplifying Substitution}
Equation~(\ref{app_rccz4_lm_theta_simp}) is not particularly well suited to 
evolution: when the lapse approaches 0, terms on the right hand side approach 
infinity. Fortunately, it can be regularized by defining a new evolutionary 
variable $\Thetat$ in terms of $\Theta$, $\alpha$, $\zh$ and $\beta^i$:
\begin{align}
   \Theta = \frac{\Thetat}{\alpha} + \frac{\beta^i \zh_i}{\alpha} .
\end{align}
In terms of these variables, we recover the evolution forms for 
$\lm\Thetat$, $\lm\Lamt^i$ and $\lm\zh_i$ expressed 
in~(\ref{rccz4_Theta2_rccz4}), (\ref{rccz4_Lh_rccz4}) and 
(\ref{rccz4_Zh_rccz4}) respectively:
\begin{align}
% verified (notes)
% verified full_cart_check_2023_09_05_
        \lm \Thetat &= \frac{\alpha^2}{2}\lp(R -\At_{ij} \At^{ij}
        +\frac{2}{3} K^2 -16\pi\rho 
        \rp.
        \\ \nonumber
        &\mathopen{}\lp.\vphantom{\frac{}{}} \hphantom{=}
        + 2\g^{ij}\Do_i \zh_j \rp) -\beta^{j}\lp( \beta^l\Do_j \zh_l + 
        \Do_j\Thetat \rp)
        \\ \nonumber
        & \hphantom{=}
        -\alpha\beta^{j} \lp( D_l {\At^{l}}_{j} -\frac{2}{3} \dt_j K 
        -8\pi j_j \rp),
\\
% verified (notes)
% verified full_car_check_2023_09_05
        \lm \Lamt^i &= \gt^{mn} \Do_m \Do_n \beta^i -2\At^{im}\dt_m\alpha
        \\ \nonumber
        & \hphantom{=}
        +2\alpha\At^{mn}\Dt^{i}\vp_{mn} +\frac{1}{3}\dt^{i}\dt_n\beta^n
        +\frac{2}{3} \Lamt^{i}\dt_n \beta^n
        \\ \nonumber
        & \hphantom{=}
        + 4\alpha \lp( \zh_j\At^{ij} 
        + 3 \At^{li}\dt_l\chi -\frac{1}{3}\dt^i K -4\pi\jt^i \rp)
        \\ \nonumber
        & \hphantom{=}
        +2\dt^i \Thetat + 2\gt^{ij}\lp( \beta^l\ \Do_{j}\zh_l 
        - \zh_l \Do_j \beta^l \rp),
\end{align}
\begin{align}
% verified (notes)
% verified full_Cart_check_2023_09_05
       \lm \zh_i &= \alpha\lp[ D_l{\At^{l}}_i -\frac{2}{3}\dt_i K 
       -8\pi j_i \rp] - \zh_l\Do_i\beta^l 
       \\ \nonumber
       & \hphantom{=}
       + \beta^{l}\Do_i \zh_{l} + \Do_i\Thetat .
%\\
\end{align}

\bibliography{rccz4}

\end{document}